\documentclass[iop,revtex4]{emulateapj}
\usepackage{lscape}

\usepackage[amssymb,mediumspace,thickqspace]{SIunits}
\addunit{\erg}{erg}
\usepackage{array}
\newcolumntype{@}{>{\global\let\currentrowstyle\relax}}
\newcolumntype{^}{>{\currentrowstyle}}
\newcommand{\rowstyle}[1]{\gdef\currentrowstyle{#1}%
  #1\ignorespaces
}

\def\arcsec{\hbox{$^{\prime\prime}$}}

\newcommand{\um}{\,\micro\metre}

\newcommand{\Fgau}{F_\nu^\mathrm{Gau}}
\newcommand{\Fpsf}{F_\nu^\mathrm{PSF}}

\newcommand{\lx}{L_\mathrm{X}}

\newcommand{\rsub}{r_\mathrm{sub}}

\newcommand{\dks}{D_\mathrm{KS}}
\newcommand{\pks}{p_\mathrm{KS}}

\newcommand{\extnuc}{R^\textrm{ext}_\textrm{nuc}}

\newcommand{\spitzerr}{{\it Spitzer}\ }                 % w/ space behind
\newcommand{\spitzer}{{\it Spitzer}}                    % w/o space behind

\newcommand{\jwstt}{{\it JWST}\ }
\newcommand{\jwst}{{\it JWST}}

\newcommand{\oiv}{[O\,IV]\ }
\newcommand{\oiii}{[O\,III]\ }

\shorttitle{Polar mid-infrared emission}
\shortauthors{Asmus et al.}

\begin{document}

 \title{The subarcsecond mid-infrared view of local active galactic nuclei:\\ III. Polar dust emission
 \thanks{Based on European Southern Observatory (ESO) observing programmes 
60.A-9242, 
074.A-9016,
075.B-0182,
075.B-0621,
 075.B-0631,
 075.B-0727, 
 075.B-0791, 
 075.B-0844, 
 076.B-0194, 
 076.B-0468, 
 076.B-0599, 
 076.B-0621, 
 076.B-0656, 
 076.B-0696, 
 076.B-0743, 
 077.B-0060, 
 077.B-0135, 
 077.B-0137, 
 077.B-0728, 
 078.B-0020, 
 078.B-0173, 
 078.B-0255, 
 078.B-0303, 
 080.B-0240, 
 080.B-0860,
 081.B-0182, 
 082.B-0299, 
 083.B-0239, 
 083.B-0452, 
 083.B-0536, 
 083.B-0592, 
 084.B-0366,
 084.B-0606, 
 084.B-0974, 
 085.B-0251, 
 085.B-0639, 
 086.B-0242, 
 086.B-0257, 
 086.B-0349, 
 086.B-0479, 
 086.B-0919, 
 087.B-0746, 
 382.A-0604, 
 382.B-0732, 
 384.B-0143, 
 384.B-0887, 
 384.B-0943, 
 385.B-0051, 
 385.B-0896, 
 385.B-0981 and 
 386.B-0026.
}
}

   \author{D.~Asmus}
   \affil{European  Southern Observatory, Casilla 19001, Santiago 19, Chile}
   \email{dasmus@eso.org}
   
	\author{S.~F.~H\"onig
          and P.~Gandhi}      
         
         \affil{Department of Physics \& Astronomy, University of Southampton, Hampshire SO17 1BJ, Southampton, United Kingdom}

\begin{abstract}
Recent mid-infrared (MIR) interferometric observations showed in few active galactic nuclei (AGN) that the bulk of the infrared emission originates from the polar region above the putative torus, where only little dust should be present.
Here, we investigate whether such strong polar dust emission is common in AGN. 
Out of 149 Seyferts in the MIR atlas of local AGN (Asmus et al.), 21 show extended MIR emission on single dish images.
In 18 objects, the extended MIR emission aligns with the system axis position angle, established by \oiii, radio, polarisation and maser based position angle measurements.
The relative amount of resolved MIR emission is at least 40 per cent and scales with the \oiv fluxes implying a strong connection between the extended continuum and \oiv emitters.  
These results together with the radio-quiet nature of the Seyferts support the scenario that the bulk of MIR emission is emitted by dust in the polar region and not by the torus, which would demand a new paradigm for the infrared emission structure in AGN.  
The current low detection rate of polar dust in the AGN of the MIR atlas is explained by the lack of sufficient high quality MIR data and the requirement for the orientation, NLR strength and distance of the AGN. 
The {\it James-Webb Space Telescope} will enable much deeper nuclear MIR studies with comparable angular resolution, allowing us to resolve the polar emission and surroundings in most of the nearby AGN.
\end{abstract}


%
%________________________________________________________________

\section{Introduction}
The nuclear infrared  emission from radio-quiet active galactic nuclei (AGN) is generally associated with dust in the parsec scale environment of the supermassive black holes \citep{antonucci_unified_1993,netzer_revisiting_2015}.
While single-telescope observations have been inefficient in directly imaging this region, infrared interferometry provided the proof of this picture over the last decade \citep[e.g.][]{jaffe_central_2004,tristram_resolving_2007,beckert_probing_2008, kishimoto_exploring_2009,kishimoto_mapping_2011, burtscher_diversity_2013}. 
In the standard unification scheme, it is assumed that the dust forms an optically- and geometrically thick ring (``dusty torus'') around the black hole, which causes many of the observed differences in the various classes of AGN.

A detailed look at the infrared emission revealed the the spectral energy distribution (SED) of the AGN contains several components with a hot-dust peak in the near-infrared and a stronger peak in the mid-infrared (MIR) being the most prominent features \citep[e.g.][]{edelson_spectral_1986,kishimoto_mapping_2011,mor_dusty_2009}.
Spectral decomposition suggested that the MIR component, which usually is the global maximum of the nuclear infrared SED, corresponds to the emission from the torus, approximately aligned with the plane of the inner accretion disk, while the weaker near-infrared peak might be hot dust in the inner narrow-line region \citep[NLR; e.g.][]{groves_infrared_2006,mor_dusty_2009,mor_hot_2012}.

Infrared interferometry currently provides the only means to directly resolve the parsec-scale dusty environment. 
Surprisingly, it was shown that the MIR, not the near-infrared, emission shows a strong polar-extended component on these small scales, accounting for more than half of the total MIR emission \citep{raban_resolving_2009,honig_parsec-scale_2012,honig_dust_2013,tristram_dusty_2014}. 
On the other hand, the near-infrared emission may originate from the disk plane \citep{honig_dust_2013}. 
If confirmed as generic features, these findings pose a major challenge to standard torus models, which assume a single emitting dusty structure confined towards the mid-plane of the AGN \citep[e.g.][]{nenkova_agn_2008,honig_dusty_2010,stalevski_3D_2012}. 

The extent of these polar wind structures can sometimes be detected on scales beyond few parsecs. 
Some nearby Seyfert galaxies display extended emission in single telescope MIR images on scales of several tens of parsecs \citep[e.g.][]{braatz_high-resolution_1993,cameron_subarcsecond_1993,bock_high_2000,radomski_high-resolution_2002,radomski_resolved_2003,whysong_thermal_2004,packham_extended_2005,reunanen_vlt_2010, honig_dusty_2010-1,asmus_subarcsecond_2014}.
While these objects preferentially show extensions towards the NLR (=polar region of the AGN), most of the detections have been serendipitous, and it is difficult to draw a more general picture of the presence of extended emission and its relation to either the torus or outflow region. 
In this work, we analyse the Seyfert galaxies from our MIR atlas of AGN observed with large telescopes from the ground \citep{asmus_subarcsecond_2014}. 
We search for extensions beyond the point spread function attributable to resolved emission of the nuclear source. 
The orientation is compared to the AGN system axis as defined by polarimetry, linear radio emission, or the NLR direction. 
A statistical analysis will allow us to conclude if the MIR emission on single-telescope scales is related to the torus or the outflow region.

\section{Data Acquisition \& Sample selection}\label{sec:sam}
The parent sample for this work is the AGN MIR atlas of 253 objects \citep{asmus_subarcsecond_2014}.
The optical classifications and distances (using the same cosmology)  are adopted from that work. 

We remove all uncertain AGN, AGN-starburst composites and low-ionisation narrow emission line region (LINER) objects from this sample to avoid non-AGN confusion and concentrate on ``classical" Seyfert objects with significant accretion rates/luminosities.
Furthermore, Mrk\,266SW and NGC\,34 are excluded because their nuclear extended MIR emission turned out to be dominated by a nuclear starburst \citep{esquej_starburst-active_2012, asmus_subarcsecond_2014}.
This cut leaves 149 MIR-detected objects, which we call the ``total Seyfert sample" in the following.

We furthermore use the nuclear MIR flux measurements from \cite{asmus_subarcsecond_2014}.
These are extracted from ground-based multi-filter photometry obtained with the instruments VLT/VISIR \citep{lagage_successful_2004}, Gemini/T-ReCS \citep{telesco_gatircam:_1998}, Gemini/Michelle \citep{glasse_michelle_1997}, and Subaru/COMICS; \citep{kataza_comics:_2000} with an angular resolution of the order of $0.35\arcsec$ or 120\,pc for the median sample distance of 72\,Mpc at 12$\,\um$.
In particular, we use the flux from Gaussian fitting, $\Fgau$, as a measure of the total nuclear flux, while the point spread function flux, $\Fpsf$, measures the unresolved flux and use measurements in various filters to compute the 12$\,\micron$ continuum fluxes as done in \cite{asmus_subarcsecond_2014}. 
The ratio of the two, $\extnuc = \Fgau / \Fpsf - 1$, will serve as a lower limit to the extended emission relative to the nuclear emission at $12\,\micron$.
   
Finally, we use the measured position angles (PAs) and extension classifications, ``point-like,'' ``unknown,'' ``possibly extended,'' and ``extended,''  from the Gaussian fitting of the nuclear emission.
In case of multiple measurements, the weighted mean of the individual PAs is used and called  ``MIR PA'' in the following.
For details of their computation and definition we refer to \cite{asmus_subarcsecond_2014}.
In summary, sources are classified as point-like if the nuclear emission remained unresolved with a point spread function close to the diffraction limit while it is classified as extended if resolved emission ($> 10$ per cent compared to point like) with consistent morphology is detected in at least two different epochs.
Otherwise the classification is possibly extended.  
All remaining cases are classified as unknown.

In order to establish system axis PAs for as many sources as possible, we consider several tracers.
First, the \oiii $\lambda 5007$ morphology shows the ionisation cones directly, unless absorption is too high. 
Thus, we search the literature for \oiii images and measured PAs (63 objects).
These are mostly based on high angular resolution data from the {\it Hubble Space Telescope} ($\sim 0.1$\,arcsec; 43 objects).
Second, a jet can be present even in radio-quiet AGN and its PA on nuclear scales is a good system axis tracer.
Therefore, we collect jet PAs from the literature using the highest angular resolution data available from very long base line interferometry, the Multi-Element Radio Linked Interferometer Network, or the Very Large Array. 
Out of 59 objects with such data, 46 have resolutions better than one arcsec and are thus comparable in scale to our MIR data (resolutions up to five arcsec for the other 7), while the frequencies are either 5 or 8.4\,GHz (or higher), except for 8 objects, where only 1.4\,GHz data are available. 
Third, if broad emission lines are detected in polarized light, their PAs are perpendicular to the system axis in case of polar scattering, as occurring in type~II objects.
Such PAs could be retrieved from the literature for 20 type~II AGN. 
Note that polarisation PAs are available also for four type~I AGN, which we do not use for our system axis estimates however, because the PAs might be dominated by equatorial scattering \citep{smith_seyferts_2004}.
Fourth and finally, five objects have resolved maser emission tracing the accretion disk. 
Thus, their PAs are as well perpendicular to the system axis.
The PA values and corresponding references of all the above tracers are given in Table~\ref{tab:sam} for the total Seyfert sample.

%\section{Methods}\label{sec:meth}

\section{Results \& Discussion}\label{sec:res}
\subsection{System axis position angle}\label{sec:SA}
None of the different system axis measurement methods is available for all of the objects.
Therefore, we combine the different methods to estimate a system axis PA whenever possible.
The median of the absolute angular difference between the \oiii and radio PAs is 8 degrees for the 31 sources which have both measurements available. 
The largest difference between both  occurs in NGC\,4278 (70\,degrees).
For 12 of the objects with measured polarisation angles in the broad emission lines and \oiii PAs, the median angular difference is 16\,degrees, with the maximum difference of 63\,degrees in NGC\,424.
The median difference between polarisation-based and radio system axis PAs is 20\,degrees from 15 objects, with the maximum differences of 63 and 65\,degrees for NGC\,6300 and NGC\,4507 respectively.
Four of the objects with maser-based system axis PAs have also \oiii PAs and the average difference between both methods is 12\,degrees. 

All these angular differences are well below 45\,degrees, i.e. the randomness limit. 
Therefore, it is justified to combine the different methods to derive one common system axis PA.
For those objects where multiple methods are available, the average of the PAs is used.
For 61 objects, none of the methods is available or delivers unambiguous results.
Their system axis PAs remain unknown.

\subsection{Alignment of system axis and mid-infrared position angles}\label{sec:align}
\begin{figure}
   \centering
   \epsscale{1.1}
   \plotone{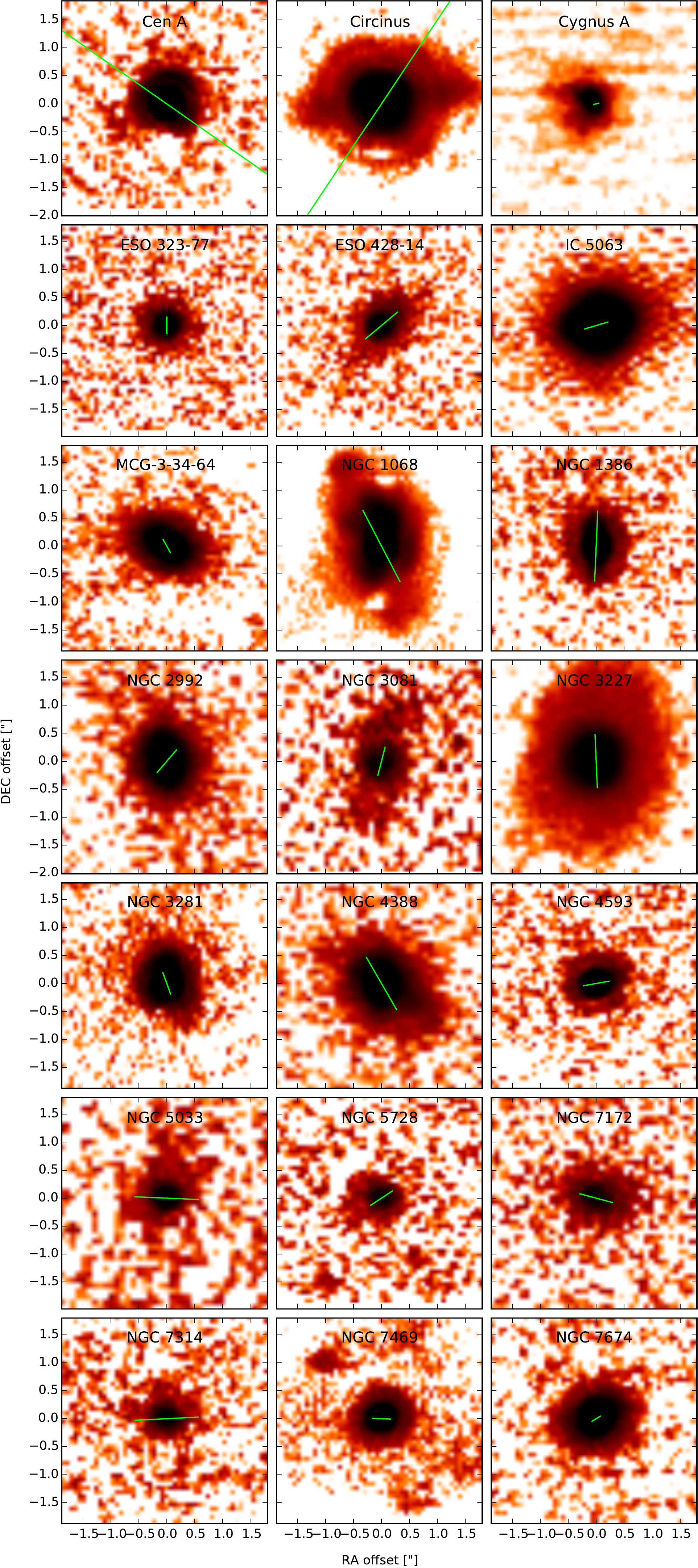}
    \caption{
Nuclear MIR emission images of the MIR-extended Seyferts from Asmus et al. (2014), except Cygnus\,A, which is from \citealt{whysong_thermal_2004}).
Filters are around 12$\,\micron$, except for NGC\,3081 ($\sim 18\,\micron$).
A square root color scale with black corresponding to 20 times the standard deviation of the background and white to the median background or lower is used. 
The green lines denote the system axis PA and is 100\,pc long. 
            }
   \label{fig:MIR_gal}
\end{figure}

In order to investigate how well the system axis and MIR PAs align, we compare the two for those objects from the total Seyfert sample that have been classified as extended in \cite{asmus_subarcsecond_2014}.  
This is the case for 22 objects. 
However, for NGC\,5506, the MIR PAs measured independently from several images are inconsistent, i.e. have a standard deviation greater than 45\,degrees. 
This leaves 21 objects which are referred to as the ``MIR-extended Seyferts" in the following.
Fig.~\ref{fig:MIR_gal} shows the nuclear MIR emission of these sources in comparison to the system axis PAs established in the previous section.

\begin{figure}
   \centering
   \epsscale{0.9}
   \plotone{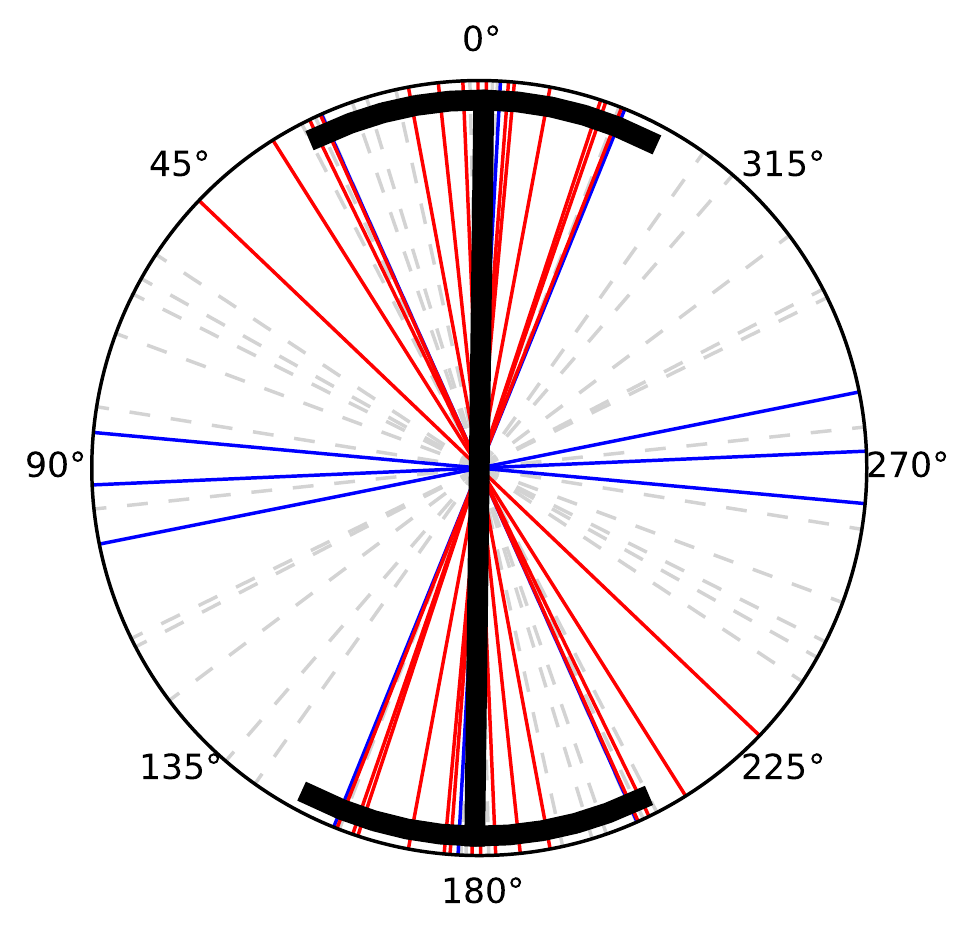}
   \epsscale{1.1}   
   \plotone{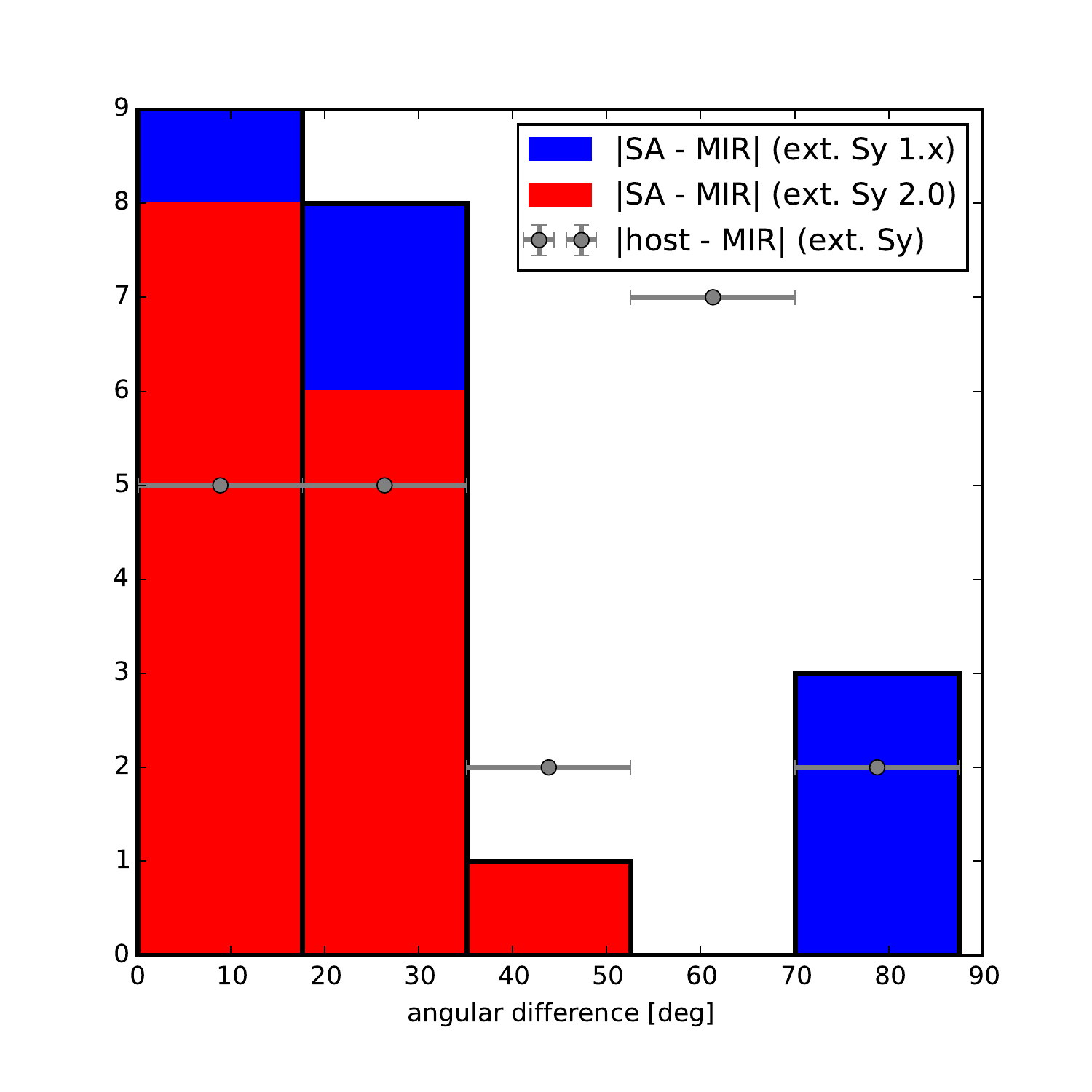}
    \caption{
             Distribution of the angular difference between the system axis and MIR PAs for the MIR-extended Seyferts.
             Top: angular plot showing individual objects as lines with Seyfert 1.x (2) in blue (red).
             The thick black line shows the mean of all extended Seyferts with the black arcs indicating the standard deviation. 
             Grey dashed lines mark the angular difference between host and MIR PAs for the same objects.
             Bottom: Additive histogram of absolute difference with the contribution of the Seyfert 1.x (2) objects is marked in blue (red).
             The distribution of the angular difference between the host and MIR PAs for all MIR-extended Seyferts is shown as grey bars. 
            }
   \label{fig:SA_MIR}
\end{figure}

The distribution of the angular difference between system axis and MIR is shown in Fig.~\ref{fig:SA_MIR}.

\begin{figure}
   \centering
   \epsscale{1.1}
   \plotone{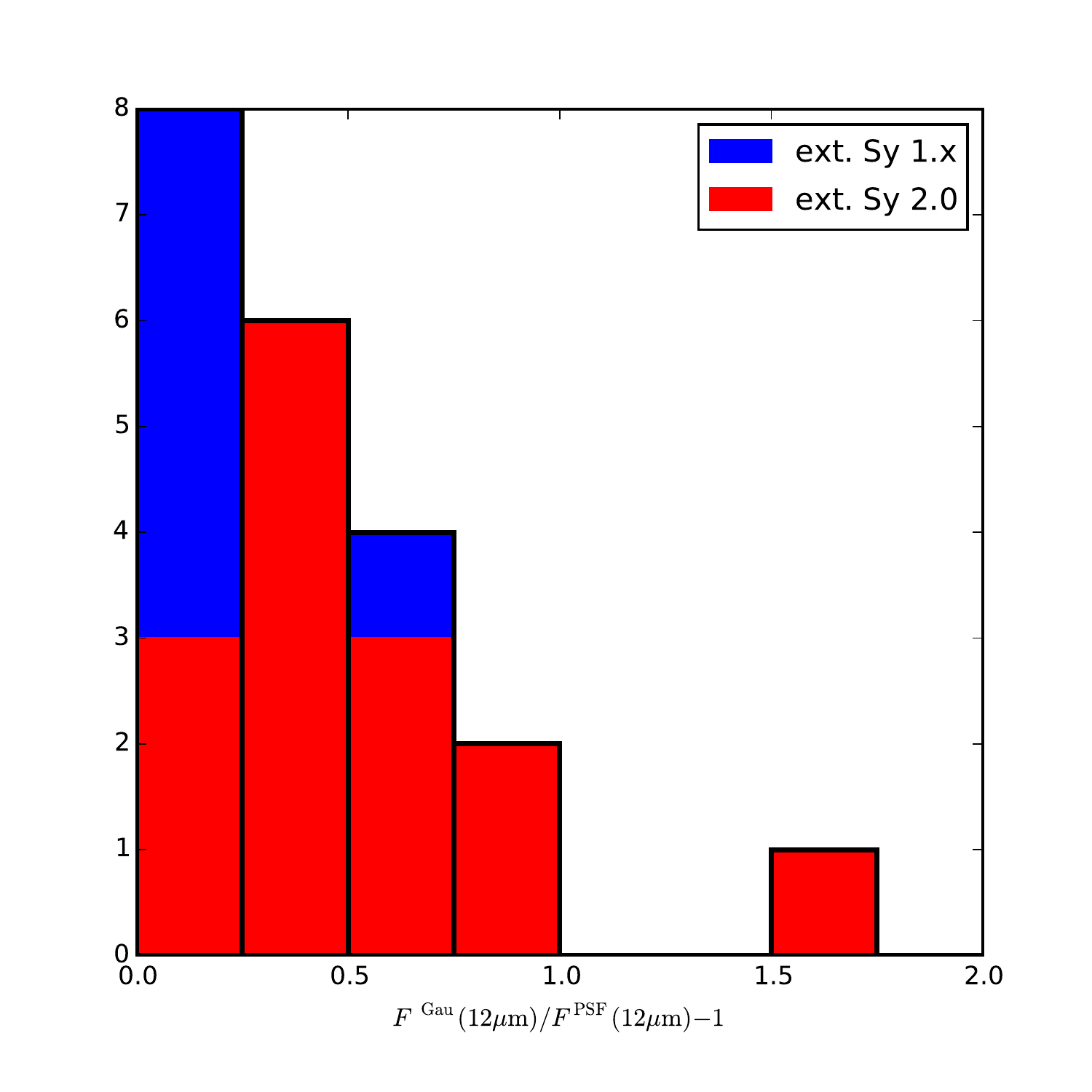}
    \caption{
             Distribution of the relative amount of extended emission with respect to the nuclear, $\extnuc$, for the MIR-extended Seyferts.
             The histogram is additive with the contribution of the Seyfert 1.x (2) objects is marked in blue (red). 
            }
   \label{fig:ext_nuc_hist}
\end{figure}
The median angular difference is 19\,degrees with a standard deviation of 27\,degrees. 
There are four objects with a difference larger than 45\,degrees (Circinus, ESO\,323-77, NGC\,2992 and NGC\,5033), which will be discussed in Sect.~\ref{sec:out1}.
Despite these, the distribution is clearly not random (corresponding to a flat line in Fig.~\ref{fig:SA_MIR}) but shows that the position angles of the nuclear MIR emission and system axis are in general connected (as already found in \citealt{asmus_subarcsecond_2014}).
We verify this result with a two-sided one-dimensional Kolmogorov-Smirnov test comparing the actual distribution to $10^5$ random uniform samples, which results in an approximate normal distribution of logarithmic null-hypothesis probabilities, $\pks$, with a median of -2.0 (standard deviation 0.9).

\subsection{Alignment of nuclear MIR emission and host structure}
The orientation and inclination of the host galaxy has been shown as independent from the system axis of the AGN (e.g., \citealt{kinney_jet_2000, schmitt_hubble_2003-1}).
We use this fact to control the  significance of the above result and compare the alignment of the nuclear MIR emission with the host structures.
For this purpose we collect the major axis position angles of the host galaxies, mainly from the Hyperleda\footnote{http://leda.univ-lyon1.fr/} data base \citep{makarov_hyperleda._2014}.
For low-inclination systems, instead of the total major axis, the PA of the inner bar or spiral structure was used. 
The corresponding value will be called ``host PA" and is listed in Table~\ref{tab:sam} for the total sample.
%The remaining 15 objects are all type~I AGN either in face-on spirals without any central structure identifiable (ESO\,511-30, Mrk\,590, and Mrk\,1239) or are very distant (12 objects).
 
The distribution of the angular difference between the MIR PA and the host PA is also shown in Fig.~\ref{fig:SA_MIR} for the MIR-extended Seyferts. 
Its median is 35\,degrees with a standard deviation of 26\,degrees. 
The shape and the median of the distribution are much closer to random than the difference between system axis and MIR PA. 
This is as well verified by a Kolmogorov-Smirnov test over $10^5$ random uniform samples with a median $\log \pks$ of just -0.27 (standard deviation 0.34).
{From this, we conclude that the extended nuclear MIR emission is indeed coming from the polar region of the AGN system rather than from the circum nuclear dust structures of the host.

\subsection{The outliers}\label{sec:out1}
There are four objects with angular differences larger 45\,degrees between the system axis and MIR PAs, which we discuss first individually here.

\subsubsection{Circinus}
Circinus has a mismatch of 46\,degrees between system axis and MIR PAs. 
The system axis PA estimate is based on \oiii images \citep{wilson_hubble_2000}, polarisation measurements \citep{oliva_spectropolarimetry_1998} and resolved maser emission \citep{greenhill_warped_2003}.
The \oiii images show a one-sided ionisation cone almost perpendicular to the maser disk while the polarisation axis is off by 23\,degrees compared to the \oiii and maser PAs.
The MIR images show bar-like emission on both sides of the nucleus \citep{asmus_subarcsecond_2014}, which aligns with the southern edge of the ionisation cone with an opening angle of roughly 90\,degrees (see also \citealt{packham_extended_2005}).
Therefore, the MIR emission is actually aligned with the system axis and indicates that the dust cone is hollow.
It remains unknown why we do not see any emission from the northern edge of the cone. 
In any case, the case of Circinus demonstrates that the mismatch between MIR and system axis PA can be up to 45 degrees and still be consistent with significant dust emission from the ionisation cone. 
Circinus is also one of the best studied object with MIR interferometry \citep{tristram_dusty_2014}, which shows that the MIR emission on sub-parsec scales is dominated by three components.
The most extended one, which contributes $\sim 80$ per cent to the total emission, has a low surface brightness and also extends along the polar axis with a similar PA as the MIR emission on larger scales.
Therefore, we conclude that Circinus is consistent with the idea that the MIR emission originates primarily from the polar region.

\subsubsection{ESO\,323-77}
The nuclear MIR emission of ESO\,323-77 is extended almost perpendicular to the system axis (85\,degrees). 
The latter is based on the \oiii morphology, which is dominated by an unresolved nucleus (typical for type~I objects) with a narrow, cone-like, possibly bended structure extending to the South for at least 5\,kpc \citep{mulchaey_emission-line_1996}.
Also MIR interferometric measurements indicate a North-South elongation on milliarcsec scales (Kishimoto et al., in prep.). 
However, the subarcsecond MIR data from \cite{asmus_subarcsecond_2014} from two epochs shows an apparent elongation close to the East-West axis although with varying position angles (standard deviation 27\,degrees). 
The marginal extent in the single dish images and the contradiction to the MIR interferometric data indicate that the measured MIR PA might be an observational artefact.

\subsubsection{NGC\,2992}
For NGC\,2992, a mismatch of 84\,degrees between MIR and system axis PAs is found.
The host is a highly inclined spiral galaxy with a prominent dust lane crossing the nucleus along the host major axis with a PA of 17 degrees \citep{ward_new_1980}.
The large biconical NLR is oriented roughly perpendicular to this (PA $\sim$ 125\,degrees; \citealt{allen_physical_1999}).
The cones are very prominent and have wide opening angles.
In the radio, the outflow is slightly more inclined to the South (PA 154\,degrees; \citealt{wehrle_radio_1988}).
In the highest signal-to-noise MIR image (N’ with Gemini/Michelle), bar like extended emission along a PA$\sim30$\,degrees is visible and consistent with morphology on the larger scale seen on the \spitzerr images \citep{asmus_subarcsecond_2014}.
The extended emission in the nuclear MIR images is extremely faint such that $\Fgau = \Fpsf$ and coincides with the dust lane.
Therefore, the dust lane dominates the extended MIR emission also on subarcsecond scales.
We speculate that the fact that no extended MIR emission from the ionization cones was detected is connected to the wide opening angle of the \oiii cones, which indicates that the system axis almost aligns with our line of sight towards the AGN.

\subsubsection{NGC\,5033}
NGC\,5033 shows the largest mismatch between MIR and system axis PAs (88\,degrees).
This type~I AGN features a one sided ionisation cone with a wide opening angle \citep{mediavilla_asymmetrical_2005}.
The jet like structure visible in the radio is aligned with the cone \citep{ho_radio_2001}.
NGC\,5033 features the weakest AGN in the MIR-extended Seyfert sample (intrinsic X-ray luminosity: $\log \lx /[\mathrm{erg}\mathrm{s}^{-1}] = 41.0$; \citealt{asmus_subarcsecond_2015}), which is one order of magnitude less powerful than Cen\,A, the second weakest.
Intense star formation dominates the MIR emission of the central few hundred parsec region as seen in the \spitzerr data \citep{asmus_subarcsecond_2014}.
The extended emission in the subarcsecond MIR images follows the host emission visible already in the \spitzer/IRAC images instead of the system axis PA.
However, \cite{asmus_subarcsecond_2014} already noted that in addition to this weak, very extended emission, the nucleus is possibly extended in East-West direction in the sharpest image (PA $\sim87$\,degrees), which would coincide with the system axis.
The signal-to-noise of the MIR images are in the lower quartile of the MIR-extended Seyfert sample and might be too low to properly detect dust emission from the polar region of this intrinsically weak AGN, in particular since the system axis is again close to our line of sight.

\subsection{Reasons for a mismatch of system axis and MIR PA}
The discussion of the individual objects above shows that an apparent mismatch of the MIR and system axis PA can have various reasons. 
On the one hand, most of the extended MIR emission can originate from the edges of the ionisation cones (as in Circinus and NGC\,1068), which is expected if the emitting material is mostly located in the walls of a hollow cone.
On the other hand, the nuclear MIR emission can be dominated by prominent host features around/in front of the nucleus like dust lanes (NGC\,2992) and intense star forming regions,  in particular if the AGN is intrinsically weak ($\log \lx/[\mathrm{erg} \mathrm{s}^{-1}] \lesssim 41.0$; NGC\,5033).
Finally, if the inclination of the system axis is low ($<45$\,degrees) as expected for type~I sources  (ESO\,323-77 and NGC\,5033) and/or the ionisation cones have wide opening angles ($>90$\,degrees; NGC\,2992, NGC\,5033), then detection of the polar dust might be difficult. 

If the above explanations are indeed correct, could at least some of the other objects in the MIR-extended Seyfert sample be affected as well?

NGC\,7172 is the only other object in this sample with a prominent dust lane crossing the nucleus. 
In fact, the PAs of the dust lane, apparent system axis, and MIR are almost perfectly aligned in this object. 
However, broad emission lines are seen in the near infrared \citep{smajic_unveiling_2012}, which implies that NGC 7172 is in fact a type~I AGN but appears as type~II only because of the dust lane.
If this is true, then it is most likely that the observed MIR extension is also caused by the dust lane.
Intense star formation in/around the nucleus is also (possibly) present in NGC\,34 and NGC\,5728.
However, both have at least one order of magnitude more powerful AGN \citep{asmus_subarcsecond_2015}, which  outshine the star formation.
The other type~I AGN among the MIR-extended Seyferts are NGC\,3227, NGC\,4593 and NGC\,7469.
Indeed, the nuclear MIR emission of NGC\,3227 shows very little elongation with significant scatter of the PA between individual images \citep{asmus_subarcsecond_2014}.
The deepest image (in the N prime filter) shows a faint halo slightly elongated in North-South direction, which matches the expectation from an almost face-on ionisation cone.
The same is true for NGC\,4593 and NGC\,7469.
Finally, none of the other objects show wide opening angles in their \oiii images.

Therefore, we conclude that the above explanations, dust lane, low-inclination or wide opening angle ionisation cone and intrinsic AGN weakness, are valid possible, but not necessarily, reasons for an observed mismatch between the system axis and MIR PAs.
The latter of these complications predominantly occur in type~I objects.
Indeed, if we look only at the 16 MIR-extended Seyfert~2 sources (red histogram in Fig.~\ref{fig:SA_MIR}), the median angular difference between system axis and MIR PA decreases to 11\,degrees (standard deviation 13\,degrees).

\subsection{Amount of extended emission and connection to \oiv emission}\label{sec:oiv}
The extent of the resolved MIR emission is difficult to determine accurately because of its low surface brightness in most of the MIR images and the unknown inclination of the system axis.  
Furthermore, an image deconvolution with the point spread function would be necessary and is beyond the scope of this work.
However, a qualitative comparison with the green lines in Fig.~\ref{fig:MIR_gal}, which were scaled to be 100\,pc long at the corresponding object distances, shows that the extent of the MIR emission varies between a few tens up to a few hundreds of parsecs.
In addition to the possibility of missing low surface brightness emission, the intrinsic resolution of the nuclear MIR images is at least ten times larger than the expected size of the torus.
For these reasons, we can only derive a lower limit for the actual amount of polar MIR emission. 
Such a lower limit is given by $\extnuc$ as described in Sect.~\ref{sec:sam}.
Its distribution for the MIR-extended Seyferts is shown in Fig.~\ref{fig:ext_nuc_hist}.
The median $\extnuc$ is 0.4 with the Seyfert~2s showing higher relative amounts of extended emission than the Seyfert~1.x objects (median 0.41 versus 0.14).

In the following, we compare $\extnuc$ to the strength of the line emission from the NLR. 
The \oiii $\lambda 5007$ line is traditionally used as a probe for the latter.
More recently, the MIR \oiv $25.89\,\micron$ emission line has been proposed as an alternative, since it is much less affected by extinction.
In addition, the line is much less affected by contamination from stellar emission due to the higher ionization potential \citep{genzel_what_1998}.
Therefore, we use the \oiv as measure of the NLR strength and collect it from the literature coming exclusively from \spitzer/IRS observations (listed in Table~\ref{tab:sam}).
%These were available for 130 objects. 

The median \oiv luminosity for the whole Seyfert sample is $1.3 \cdot 10^{41}$\,erg s$^{-1}$, while the corresponding one of the extended sources is slightly higher ($2.0 \cdot 10^{41}$\,erg\,s$^{-1}$\,cm$^{-2}$).
Furthermore, the median flux is seven times higher ($9.9 \cdot 10^{-13}$ versus $1.4 \cdot 10^{-13}$\,erg\,s$^{-1}$\,cm$^{-2}$).
This separation in \oiv flux becomes even clearer if one compares the extended objects to the 34 point-like Seyferts from the total sample only, as shown in Fig.~\ref{fig:oiv_hist}.
\begin{figure}
   \centering
      \epsscale{1.1}
   \plotone{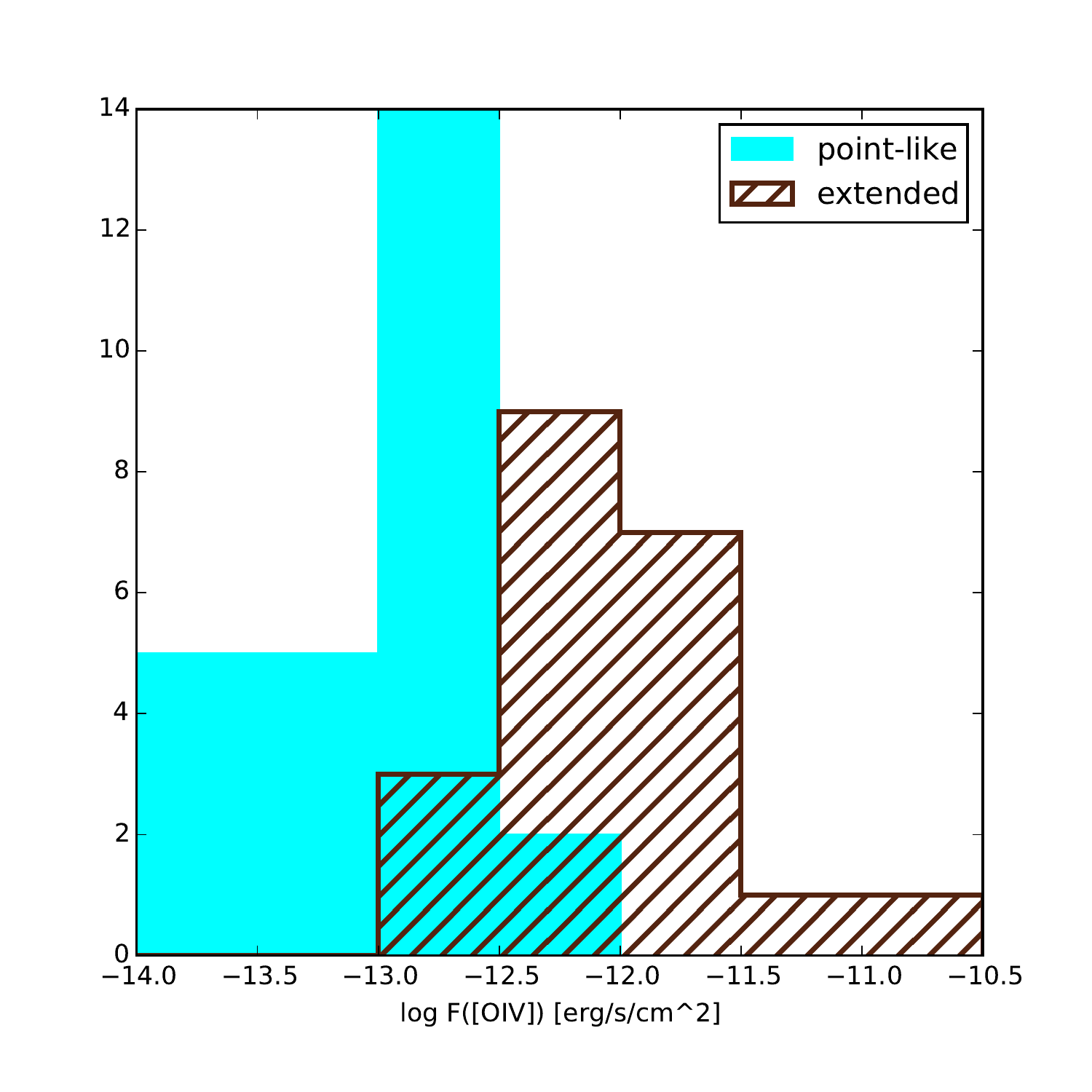}
    \caption{
             \oiv flux distribution for the point-like Seyferts (cyan filled histogram) versus the MIR-extended Seyferts (brown hatched histogram). 
            }
   \label{fig:oiv_hist}
\end{figure}
The flux distributions of both groups barely overlap (ratio of medians is 7.9, Kolmogorov-Smirnov test distance is 0.82 with $\log \pks =  -7.1$). 
The separation in \oiv flux is by far the strongest among several other parameters like nuclear MIR flux (ratio of medians: 1.4; fluxes from \citealt{asmus_subarcsecond_2014}), intrinsic 2-10\,keV flux (ratio of medians: 1.7; fluxes from \citealt{asmus_subarcsecond_2015}) and distance (ratio of medians: 3.0).
In particular, there is no significant separation between extended and point-like sources when using  \oiii fluxes instead of \oiv (ratio of medians: 1.5, $\dks = 0.23$; $\log \pks = -0.21$; mainly taken from \citealt{ho_search_1997,gu_emission-line_2006,diamond-stanic_isotropic_2009}), which demonstrates why we prefer \oiv over \oiii emission. 
This might be related to recent results questioning the \oiii emission as accurate bolometric indicator \citep{berney_bat_2015,ueda_[o_2015}.

If we compare the amount of extended  emission, $\extnuc$, to the \oiv flux, in Fig.~\ref{fig:oiv_ext_nuc}, indeed, they correlate well.
\begin{figure}
   \centering
   \epsscale{1.2}
   \plotone{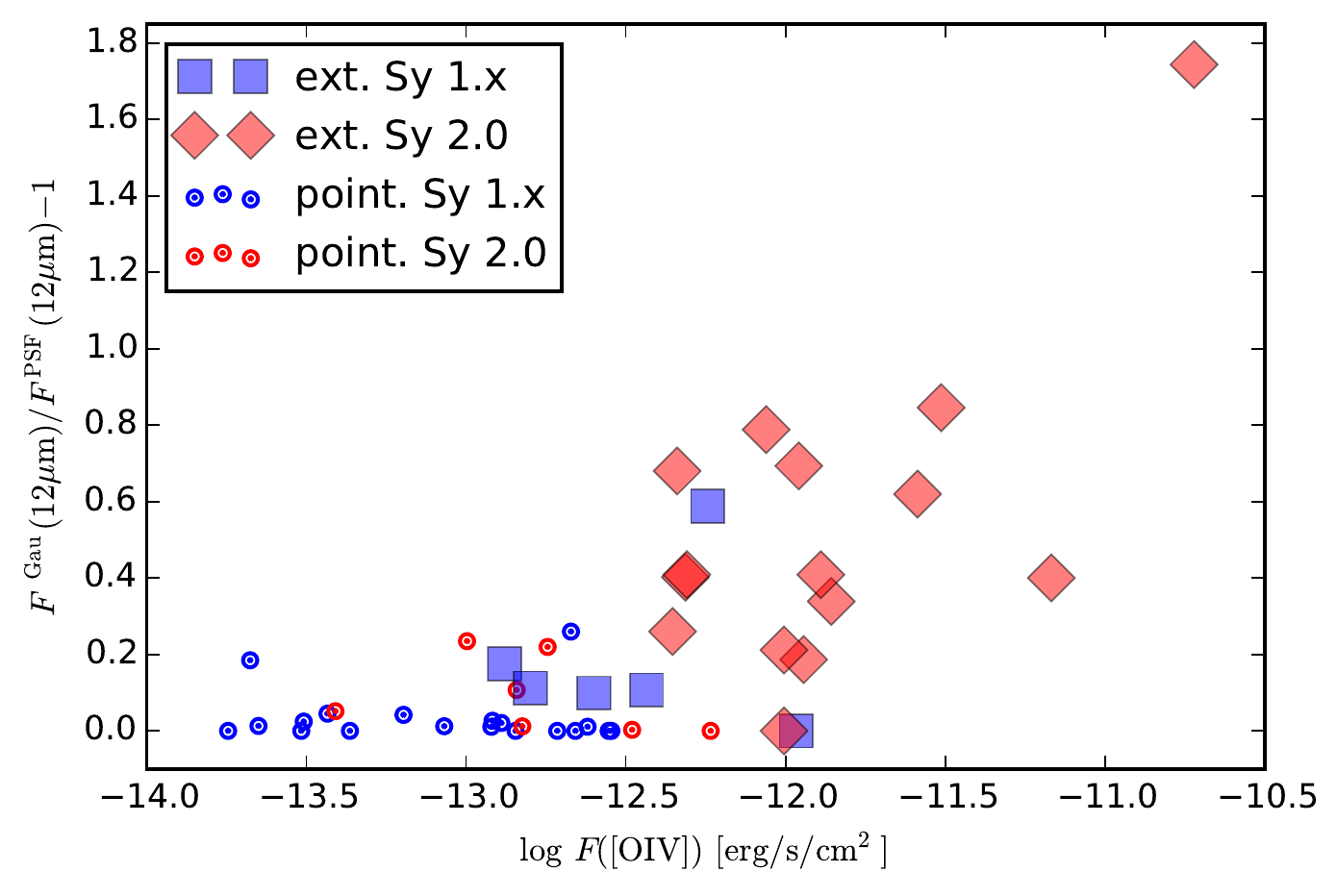}
    \caption{
    Relative amount of extended emission with respect to the nuclear, $\extnuc$, versus the \oiv flux for the MIR-extended Seyferts.
           Seyfert~2 objects are marked by red diamonds while Seyfert~1.x are blue squares. 
           For comparison, the point-like Seyferts from the total sample are shown with $\odot$ symbols (blue: Sy~1.x; red: Sy~2).
            }
   \label{fig:oiv_ext_nuc}
\end{figure}
In logarithmic space, the Spearman rank is 0.54 with a null-hypothesis probability of 0.012. 
For comparison, we again show the point-like Seyferts from the total sample, which all have low $\extnuc$ values as expected.
Interestingly, the few objects in the overlapping region are mostly Seyfert~2s in the point-like sample and Seyfert~1.x in the extended sample.
Extended emission appears to be easier detectable in Seyfert~2s, as we expect from the side-ways view onto the ionization cones (see also next section).

These results imply that the extended MIR continuum emission is strongly connected to the \oiv emission, e.g., coming from the same region.
Furthermore, the object distance also plays an important role in the detection rate of extended MIR emission. 
This will be further discussed in the next section.

The presence of dust in the narrow emission line clouds was already inferred by \cite{netzer_dust_1993} and recently was further investigated by \cite{stern_radiation_2014}. 
This work examines the structure of ionised gas clouds, e.g. as seen in the NLR, and find that they are capable of hosting dust (high density, intrinsic shielding), when considering that the cloud structure is set by the radiation pressure (“radiation pressure confinement”). 
As such, line and dust emission originate from the same clouds. 
Indeed, the authors predict a relation between cloud density and luminosity of the ionising part of the AGN emission. 
Our finding lends support to this idea by linking the dust emission strength and its place of origin to the strength of a highly ionised emission line.

\subsection{Reasons for low detection rate of polar emission}
The results of the previous section indicate that the object distance is an important parameter for the detection of extended MIR emission. 
In fact, the MIR-extended Seyferts are less distant from us than most Seyferts in the total sample (median distance 40\,Mpc compared to 94\,Mpc of the total sample), allowing to reach intrinsic resolutions of less than 100\,pc. 
Still, in most of the MIR-extended Seyferts, the polar emission is resolved with full-width-half-maxima less than two times the point spread function, which makes it difficult to separate the MIR emission into different components. 
 
Another reason for the low detection rate is the relatively low signal-to-noise of most of the MIR images in \cite{asmus_subarcsecond_2014}. 
The median peak signal-to-noise of the detected nuclei is 25, while it is 93 for the MIR-extended Seyferts.
And, the extended emission has a low surface brightness in most of the cases (exceptions are NGC\,1068 and NGC\,1386).

The median ratio of unresolved to total emission for the extended Seyfert 2s is 0.7 (standard deviation 0.2), which provides a typical lower limit of 30 per cent for the polar emission on scales resolvable with single 8\,meter class telescopes.
Furthermore, the example of NGC\,5033 indicates that the AGN also has to be sufficiently powerful to create sufficient polar dust emission ($\log \lx/[\mathrm{erg} \mathrm{s}^{-1}] \gtrsim 42$).
Finally, we showed that for type~I sources, the polar dust is difficult to detect because of the low inclination and projected wide opening angle of the ionisation cones. 
%Finally, one can imagine that for various reasons the ionisation cones are intrinsically small or weakly emitting in some objects, which would also complicate the their detection in the mid-infrared. 

From the above reasoning one can form a set of selection criteria to find the best candidates among the 128 Seyferts without any detections of polar MIR emission so far to look for the latter.
We base the criteria on the MIR-extended sample and start with a peak signal-to-noise of the nuclear MIR source greater than 36 (first quartile), which leaves 45 objects.
The effects of distance, AGN power and NLR strength can be combined into one criterion based on the \oiv flux, which should be greater than $4.6 \cdot 10^{-13}$\,erg\,s$^{-1}$\,cm$^{-2}$ (first quartile). 
These two criteria leave eight sources that should have detectable extended MIR emission.
Although four among them are not Seyfert 2s, two of these, IC\,4329A and NGC\,4151, show extensions in the archival nuclear MIR images consistent with the system axis orientation (see also \citealt{radomski_resolved_2003}), while the other two type~I AGN, 3C\,120 and NGC\,1365, exhibit as well extended emission although with inconsistent MIR PAs between different epochs.
Such inconsistencies can arise from instabilities in the integrated point spread function in case of technical imperfections or poor seeing conditions.  
The Seyfert~2s among the eight candidates are IC\,4518W, Mrk\,573, NGC\,5135 and NGC\,5506.
IC\,4518W is classified as possibly extended but with no system axis PA information available, while Mrk\,573 is also possibly extended with a MIR PA consistent with the system axis PA, but lacks a suitable standard star observation for a robust morphological classification. 
NGC\,5506 was excluded earlier from the extended Seyfert sample because of inconsistent MIR PA measurements, leaving NGC\,5135 as the only source which shows no extension among these best candidates.

In summary, a careful revaluation of the MIR data indicates that polar MIR emission might indeed be present in at least seven to eight further objects, beside the 18 objects, where we found such features outright. 
Additional deep high-quality MIR imaging with subarcsecond resolution might reveal the polar emission in these promising candidates.

\section{Conclusions \& Implications}\label{sec:imp}
Of the 149 MIR detected Seyferts from the AGN MIR atlas, 18 have clear detections of extended nuclear MIR emission, which comes from the polar region of the AGN on scales of tens to hundreds of parsecs, i.e., much larger scales than expected torus sizes. 
Owing to the fact that only three of these sources are radio-loud (Cen\,A, Cygnus\,A and IC\,5063), synchrotron emission from the jet can in general not be the main contributor for this emission.
In addition, the strong connection to the \oiv emission implies that the extended MIR emission originates in the NLR.
The best candidate for the emitter is thus dust in the walls of the ionization cone and/or NLR clouds.

The extended emission that we could detect with single-dish observations represents only a lower limit of the total polar emission.
In fact, all objects with detailed MIR interferometric observations have shown that the large majority of MIR emission is coming from an extended polar component, which presumably is the same we see on much larger scales in this work. 
This argues for a modification of the current paradigm in which the bulk of the mid-infrared emission is emitted by the canonical parsec-scale torus.
The latter conclusively might be much more compact or even just a thick part of the accretion disk.
In both cases, its dust would then be responsible for the hot component in the infrared SED.

The alignment of the polar MIR emission with the edges of the ionization cones (as in Circinus) indicates that the polar MIR emitter is located rather in the walls of the cone, which thus would remain rather hollow. 
This scenario will presumably have consequences on, e.g., the absorption geometry, kinematics and resulting SED, whose study we leave for future work. 
However, unless the walls are optically thick, which is unlikely as we see the \oiii from the inside, we expect the MIR emission to be even more isotropic than in the case of the canonical clumpy torus model, which can naturally account for the lack of differences between type~I and II AGN in the MIR--X-ray correlation \citep{krabbe_n-band_2001,gandhi_resolving_2009,asmus_subarcsecond_2015}.

The relatively low detection rate of polar emission in AGN so far can be explained by the combination of the lack of sufficient high quality data and the requirement for the orientation, NLR strength and distance of the AGN.

Looking forward to the future, the upgraded VISIR instrument \citep{kaufl_return_2015} allows us to refine the extension analysis for the brighter of the currently ambiguous cases in particular with the burst mode, optimizing image quality and sensitivity
In the near future, MATISSE \citep{lopez_overview_2014} will provide improved interferometric capabilities to characterise the milliarcsecond scale MIR emission of the two dozen brightest AGN.
Larger samples will only be accessible with the {\it James-Webb Space Telescope} (\jwst; \citealt{gardner_james_2006}), which is expected to provide a 10$^3$ times higher sensitivity at 12$\um$ than current ground-based MIR instruments, while offering a comparable angular resolution.
Thus, deep \jwstt imaging will allow for the detection and resolution of the polar emission in most nearby Seyferts, in particular in type~II objects with at least moderate  \oiv luminosities.
The size of the polar dust emission is expected to scale with the bolometric luminosity of the AGN such that even more distant systems can be resolved, provided they are powerful enough. 
The most distant AGN in our resolved Seyfert sample is Cygnus\,A at 257\,Mpc with a bolometric luminosity of $\sim 10^{45}$\,erg\,s$^{-1}$. 
The distance versus luminosity threshold for the detection of polar emission can be expressed by the intrinsic resolution in units of the sublimation radius, $\rsub = 0.47 \sqrt{ L_\mathrm{bol} / 10^{46}\mathrm{erg}\,\mathrm{s}^{-1}}$\,pc \citep{kishimoto_innermost_2007}.
Here, the extended Seyferts exhibit elongations up to $\sim10^3 \rsub$ (maximum is NGC\,5033).
Thus, \jwstt should be able to resolve polar emission in AGN up to similar values.
Of course, the detection will depend on the actual NLR size and emission power in the individual cases, which shows some scatter with respect to the bolometric luminosity. 
The \oiv flux may be the main selection criterion here.

%Furthermore, the extent of the polar emission will possibly allow for putting constraints on the inclination of the AGN system axis, in particular in combination with data from other wavelengths, like \oiii images. 
%The system axis inclination is an important unknown, e.g., in the modelling of the obscuration in individual objects.
Finally, the deep MIR images from \jwstt will also show the transition zone of the AGN to the host structure and thus help to separate foreground from AGN-intrinsic obscuration.

In the more distant future, the Planet Finding Imager \citep{kraus_science_2014} will provide 0.1\,mas resolution images at $\sim 12\,\micron$, almost 3000\,$\times$ better than current single telescope observations, and thus allow detailed investigations of inner parts of nearby, bright AGN.

\acknowledgments
We thank the anonymous referee for the helpful comments leading to an improved paper.
We also thank David Whysong for providing the Keck/LWS image of Cygnus\,A.
DA thanks Marko Stalevski for fruitful discussion and comments.
SFH acknowledges support from the Marie Curie International Incoming Fellowship within the 7th European Community Framework Programme (PIIF- GA-2013-623804).
PG acknowledges support from STFC (grant reference ST/J003697/1).
We acknowledge the usage of the HyperLeda database (http://leda.univ-lyon1.fr).
This research made use of the NASA/IPAC Extragalactic Database
(NED), which is operated by the Jet Propulsion Laboratory, California Institute
of Technology, under contract with the National Aeronautics and Space
Administration.

% 
% for the bibliography, at the end
%\bibliographystyle{aa} % style aa.bst
\bibliographystyle{mn2e}

\bibliography{my_lib_ref.bib} % your references Yourfile.bib

\begin{thebibliography}{182}
\expandafter\ifx\csname natexlab\endcsname\relax\def\natexlab#1{#1}\fi

\bibitem[{Allen {et~al}\mbox{.}(1999)Allen, Dopita, Tsvetanov, \&
  Sutherland}]{allen_physical_1999}
Allen M.~G., Dopita M.~A., Tsvetanov Z.~I., Sutherland R.~S., 1999, ApJ, 511,
  686

\bibitem[{Antonucci(1993)}]{antonucci_unified_1993}
Antonucci R., 1993, ARA\&A, 31, 473

\bibitem[{Antonucci(1985)}]{antonucci_vla_1985}
Antonucci R. R.~J., 1985, ApJSS, 59, 499

\bibitem[{Asmus {et~al}\mbox{.}(2015)Asmus, Gandhi, H\"onig, Smette, \&
  Duschl}]{asmus_subarcsecond_2015}
Asmus D., Gandhi P., H\"onig S.~F., Smette A., Duschl W.~J., 2015, MNRAS, 454,
  766

\bibitem[{Asmus {et~al}\mbox{.}(2014)Asmus, H\"onig, Gandhi, Smette, \&
  Duschl}]{asmus_subarcsecond_2014}
Asmus D., H\"onig S.~F., Gandhi P., Smette A., Duschl W.~J., 2014, MNRAS, 439,
  1648

\bibitem[{Bailey {et~al}\mbox{.}(1988)Bailey, Axon, Hough, Ward, McLean, \&
  Heathcote}]{bailey_polarization_1988}
Bailey J., Axon D.~J., Hough J.~H., Ward M.~J., McLean I., Heathcote S.~R.,
  1988, MNRAS, 234, 899

\bibitem[{Baum {et~al}\mbox{.}(1988)Baum, Heckman, Bridle, van Breugel, \&
  Miley}]{baum_extended_1988}
Baum S.~A., Heckman T.~M., Bridle A., van Breugel W. J.~M., Miley G.~K., 1988,
  ApJSS, 68, 643

\bibitem[{Beckert {et~al}\mbox{.}(2008)Beckert, Driebe, H\"onig, \&
  Weigelt}]{beckert_probing_2008}
Beckert T., Driebe T., H\"onig S.~F., Weigelt G., 2008, A\&A, 486, L17

\bibitem[{Bennert {et~al}\mbox{.}(2002)Bennert, Falcke, Schulz, Wilson, \&
  Wills}]{bennert_size_2002}
Bennert N., Falcke H., Schulz H., Wilson A.~S., Wills B.~J., 2002, The ApJL,
  574, L105

\bibitem[{Bennert {et~al}\mbox{.}(2006)Bennert, Jungwiert, Komossa, Haas, \&
  Chini}]{bennert_size_2006}
Bennert N., Jungwiert B., Komossa S., Haas M., Chini R., 2006, A\&A, 446, 919

\bibitem[{Berney {et~al}\mbox{.}(2015)Berney, Koss, Trakhtenbrot, Ricci,
  Lamperti, Schawinski, Balokovi\'c, Crenshaw, Fischer, Gehrels, Harrison,
  Hashimoto, Ichikawa, Mushotzky, Oh, Stern, Treister, Ueda, Veilleux, \&
  Winter}]{berney_bat_2015}
Berney S. {et~al.}, 2015, MNRAS, 454, 3622

\bibitem[{Black {et~al}\mbox{.}(1992)Black, Baum, Leahy, Perley, Riley, \&
  Scheuer}]{black_study_1992}
Black A. R.~S., Baum S.~A., Leahy J.~P., Perley R.~A., Riley J.~M., Scheuer P.
  A.~G., 1992, MNRAS, 256, 186

\bibitem[{Bock {et~al}\mbox{.}(2000)Bock, Neugebauer, Matthews, Soifer,
  Becklin, Ressler, Marsh, Werner, Egami, \& Blandford}]{bock_high_2000}
Bock J.~J. {et~al.}, 2000, AJ, 120, 2904

\bibitem[{Boyce {et~al}\mbox{.}(1996)Boyce, Disney, Macchetto, Boksenberg,
  Blades, \& Mackay}]{boyce_faint_1996}
Boyce P.~J., Disney M.~J., Macchetto F., Boksenberg A., Blades J.~C., Mackay
  C.~D., 1996, A\&A, 305, 715

\bibitem[{Braatz {et~al}\mbox{.}(1993)Braatz, Wilson, Gezari, Varosi, \&
  Beichman}]{braatz_high-resolution_1993}
Braatz J.~A., Wilson A.~S., Gezari D.~Y., Varosi F., Beichman C.~A., 1993, The
  ApJL, 409, L5

\bibitem[{Burtscher {et~al}\mbox{.}(2013)Burtscher, Meisenheimer, Tristram,
  Jaffe, H\"onig, Davies, Kishimoto, Pott, R\"ottgering, Schartmann, Weigelt,
  \& Wolf}]{burtscher_diversity_2013}
Burtscher L. {et~al.}, 2013, A\&A, 558, 149

\bibitem[{Cameron {et~al}\mbox{.}(1993)Cameron, Storey, Rotaciuc, Genzel,
  Verstraete, Drapatz, Siebenmorgen, \& Lee}]{cameron_subarcsecond_1993}
Cameron M., Storey J. W.~V., Rotaciuc V., Genzel R., Verstraete L., Drapatz S.,
  Siebenmorgen R., Lee T.~J., 1993, ApJ, 419, 136

\bibitem[{Canalizo {et~al}\mbox{.}(2003)Canalizo, Max, Whysong, Antonucci, \&
  Dahm}]{canalizo_adaptive_2003}
Canalizo G., Max C., Whysong D., Antonucci R., Dahm S.~E., 2003, ApJ, 597, 823

\bibitem[{Capetti {et~al}\mbox{.}(1995)Capetti, Macchetto, Axon, Sparks, \&
  Boksenberg}]{capetti_morphology_1995}
Capetti A., Macchetto F., Axon D.~J., Sparks W.~B., Boksenberg A., 1995, ApJ,
  448, 600

\bibitem[{Claussen, Heiligman \& Lo(1984)Claussen, Heiligman, \&
  Lo}]{claussen_water-vapor_1984}
Claussen M.~J., Heiligman G.~M., Lo K.~Y., 1984, Nature, 310, 298

\bibitem[{Cohen {et~al}\mbox{.}(1999)Cohen, Ogle, Tran, Goodrich, \&
  Miller}]{cohen_polarimetry_1999}
Cohen M.~H., Ogle P.~M., Tran H.~D., Goodrich R.~W., Miller J.~S., 1999, AJ,
  118, 1963

\bibitem[{Colbert {et~al}\mbox{.}(1996)Colbert, Baum, Gallimore, O'Dea,
  Lehnert, Tsvetanov, Mulchaey, \& Caganoff}]{colbert_large-scale_1996}
Colbert E. J.~M., Baum S.~A., Gallimore J.~F., O'Dea C.~P., Lehnert M.~D.,
  Tsvetanov Z.~I., Mulchaey J.~S., Caganoff S., 1996, ApJSS, 105, 75

\bibitem[{Cooke {et~al}\mbox{.}(2000)Cooke, Baldwin, Ferland, Netzer, \&
  Wilson}]{cooke_narrow-line_2000}
Cooke A.~J., Baldwin J.~A., Ferland G.~J., Netzer H., Wilson A.~S., 2000,
  ApJSS, 129, 517

\bibitem[{Cracco {et~al}\mbox{.}(2011)Cracco, Ciroi, di~Mille, Vaona, Frassati,
  Smirnova, La~Mura, Moiseev, \& Rafanelli}]{cracco_origin_2011}
Cracco V. {et~al.}, 2011, MNRAS, 418, 2630

\bibitem[{Dasyra {et~al}\mbox{.}(2008)Dasyra, Ho, Armus, Ogle, Helou, Peterson,
  Lutz, Netzer, \& Sturm}]{dasyra_high-ionization_2008}
Dasyra K.~M. {et~al.}, 2008, ApJ, 674, L9

\bibitem[{Dasyra {et~al}\mbox{.}(2011)Dasyra, Ho, Netzer, Combes, Trakhtenbrot,
  Sturm, Armus, \& Elbaz}]{dasyra_view_2011}
Dasyra K.~M., Ho L.~C., Netzer H., Combes F., Trakhtenbrot B., Sturm E., Armus
  L., Elbaz D., 2011, ApJ, 740, 94

\bibitem[{Diamond-Stanic, Rieke \& Rigby(2009)Diamond-Stanic, Rieke, \&
  Rigby}]{diamond-stanic_isotropic_2009}
Diamond-Stanic A.~M., Rieke G.~H., Rigby J.~R., 2009, ApJ, 698, 623

\bibitem[{Dicken {et~al}\mbox{.}(2014)Dicken, Tadhunter, Morganti, Axon,
  Robinson, Magagnoli, Kharb, {Ramos Almeida}, Mingo, Hardcastle, Nesvadba,
  Singh, Kouwenhoven, Rose, Spoon, Inskip, \& Holt}]{dicken_spitzer_2014}
Dicken D. {et~al.}, 2014, ApJ, 788, 98

\bibitem[{Durret \& Bergeron(1987)}]{durret_imaging_1987}
Durret F., Bergeron J., 1987, A\&A, 173, 219

\bibitem[{Durret \& Bergeron(1988)}]{durret_long_1988}
Durret F., Bergeron J., 1988, A\&A Supplement Series, 75, 273

\bibitem[{Edelson \& Malkan(1986)}]{edelson_spectral_1986}
Edelson R.~A., Malkan M.~A., 1986, ApJ, 308, 59

\bibitem[{Erwin(2004)}]{erwin_double-barred_2004}
Erwin P., 2004, A\&A, 415, 941

\bibitem[{Esquej {et~al}\mbox{.}(2012)Esquej, Alonso-Herrero, P\'erez-Garc\'ia,
  Pereira-Santaella, Rigopoulou, S\'anchez-Portal, Castillo, {Ramos Almeida},
  Coia, Altieri, Acosta-Pulido, Conversi, Gonz\'alez-Serrano, Hatziminaoglou,
  Povi\'c, Rodr\'iguez-Espinosa, \& Valtchanov}]{esquej_starburst-active_2012}
Esquej P. {et~al.}, 2012, MNRAS, 423, 185

\bibitem[{Evans {et~al}\mbox{.}(1991)Evans, Ford, Kinney, Antonucci, Armus, \&
  Caganoff}]{evans_hst_1991}
Evans I.~N., Ford H.~C., Kinney A.~L., Antonucci R. R.~J., Armus L., Caganoff
  S., 1991, The ApJL, 369, L27

\bibitem[{Evans {et~al}\mbox{.}(1993)Evans, Tsvetanov, Kriss, Ford, Caganoff,
  \& Koratkar}]{evans_hubble_1993}
Evans I.~N., Tsvetanov Z., Kriss G.~A., Ford H.~C., Caganoff S., Koratkar
  A.~P., 1993, ApJ, 417, 82

\bibitem[{Falcke, Wilson \& Simpson(1998)Falcke, Wilson, \&
  Simpson}]{falcke_hst_1998}
Falcke H., Wilson A.~S., Simpson C., 1998, ApJ, 502, 199

\bibitem[{Falcke {et~al}\mbox{.}(1996)Falcke, Wilson, Simpson, \&
  Bower}]{falcke_helical_1996}
Falcke H., Wilson A.~S., Simpson C., Bower G.~A., 1996, The ApJL, 470, L31

\bibitem[{Farrah {et~al}\mbox{.}(2007)Farrah, Bernard-Salas, Spoon, Soifer,
  Armus, Brandl, Charmandaris, Desai, Higdon, Devost, \&
  Houck}]{farrah_high-resolution_2007}
Farrah D. {et~al.}, 2007, ApJ, 667, 149

\bibitem[{Ferruit, Wilson \& Mulchaey(2000)Ferruit, Wilson, \&
  Mulchaey}]{ferruit_hubble_2000}
Ferruit P., Wilson A.~S., Mulchaey J., 2000, ApJSS, 128, 139

\bibitem[{Filippenko \& Ho(2003)}]{filippenko_low-mass_2003}
Filippenko A.~V., Ho L.~C., 2003, The ApJL, 588, L13

\bibitem[{Ford {et~al}\mbox{.}(1985)Ford, Crane, Jacoby, Lawrie, \& van~der
  Hulst}]{ford_bubbles_1985}
Ford H.~C., Crane P.~C., Jacoby G.~H., Lawrie D.~G., van~der Hulst J.~M., 1985,
  ApJ, 293, 132

\bibitem[{Fraquelli, Storchi-Bergmann \& Binette(2000)Fraquelli,
  Storchi-Bergmann, \& Binette}]{fraquelli_extended_2000}
Fraquelli H.~A., Storchi-Bergmann T., Binette L., 2000, ApJ, 532, 867

\bibitem[{Gallimore {et~al}\mbox{.}(1996{\natexlab{a}})Gallimore, Baum, O'Dea,
  Brinks, \& Pedlar}]{gallimore_h_1996}
Gallimore J.~F., Baum S.~A., O'Dea C.~P., Brinks E., Pedlar A.,
  1996{\natexlab{a}}, ApJ, 462, 740

\bibitem[{Gallimore {et~al}\mbox{.}(1996{\natexlab{b}})Gallimore, Baum, O'Dea,
  \& Pedlar}]{gallimore_subarcsecond_1996-1}
Gallimore J.~F., Baum S.~A., O'Dea C.~P., Pedlar A., 1996{\natexlab{b}}, ApJ,
  458, 136

\bibitem[{Gandhi {et~al}\mbox{.}(2009)Gandhi, Horst, Smette, H\"onig, Comastri,
  Gilli, Vignali, \& Duschl}]{gandhi_resolving_2009}
Gandhi P., Horst H., Smette A., H\"onig S., Comastri A., Gilli R., Vignali C.,
  Duschl W., 2009, A\&A, 502, 457

\bibitem[{Gardner {et~al}\mbox{.}(2006)Gardner, Mather, Clampin, Doyon,
  Greenhouse, Hammel, Hutchings, Jakobsen, Lilly, Long, Lunine, McCaughrean,
  Mountain, Nella, Rieke, Rieke, Rix, Smith, Sonneborn, Stiavelli, Stockman,
  Windhorst, \& Wright}]{gardner_james_2006}
Gardner J.~P. {et~al.}, 2006, Space Science Reviews, 123, 485

\bibitem[{Genzel {et~al}\mbox{.}(1998)Genzel, Lutz, Sturm, Egami, Kunze,
  Moorwood, Rigopoulou, Spoon, Sternberg, Tacconi-Garman, Tacconi, \&
  Thatte}]{genzel_what_1998}
Genzel R. {et~al.}, 1998, ApJ, 498, 579

\bibitem[{Giovannini {et~al}\mbox{.}(2005)Giovannini, Taylor, Feretti, Cotton,
  Lara, \& Venturi}]{giovannini_bologna_2005}
Giovannini G., Taylor G.~B., Feretti L., Cotton W.~D., Lara L., Venturi T.,
  2005, ApJ, 618, 635

\bibitem[{Glasse, Atad-Ettedgui \& Harris(1997)Glasse, Atad-Ettedgui, \&
  Harris}]{glasse_michelle_1997}
Glasse A.~C., Atad-Ettedgui E.~I., Harris J.~W., 1997, in {SPIE}, Vol. 2871,
  Optical {Telescopes} of {Today} and {Tomorrow}, pp. 1197--1203

\bibitem[{Gonz\'alez~Delgado {et~al}\mbox{.}(1998)Gonz\'alez~Delgado, Heckman,
  Leitherer, Meurer, Krolik, Wilson, Kinney, \&
  Koratkar}]{gonzalez_delgado_ultraviolet-optical_1998}
Gonz\'alez~Delgado R.~M., Heckman T., Leitherer C., Meurer G., Krolik J.,
  Wilson A.~S., Kinney A., Koratkar A., 1998, ApJ, 505, 174

\bibitem[{Greenhill {et~al}\mbox{.}(2003)Greenhill, Booth, Ellingsen,
  Herrnstein, Jauncey, McCulloch, Moran, Norris, Reynolds, \&
  Tzioumis}]{greenhill_warped_2003}
Greenhill L.~J. {et~al.}, 2003, ApJ, 590, 162

\bibitem[{Greenhill {et~al}\mbox{.}(1996)Greenhill, Gwinn, Antonucci, \&
  Barvainis}]{greenhill_vlbi_1996}
Greenhill L.~J., Gwinn C.~R., Antonucci R., Barvainis R., 1996, The ApJL, 472,
  L21

\bibitem[{Greenhill {et~al}\mbox{.}(1995)Greenhill, Jiang, Moran, Reid, Lo, \&
  Claussen}]{greenhill_detection_1995}
Greenhill L.~J., Jiang D.~R., Moran J.~M., Reid M.~J., Lo K.~Y., Claussen
  M.~J., 1995, ApJ, 440, 619

\bibitem[{Groves, Dopita \& Sutherland(2006)Groves, Dopita, \&
  Sutherland}]{groves_infrared_2006}
Groves B., Dopita M., Sutherland R., 2006, A\&A, 458, 405

\bibitem[{Gu {et~al}\mbox{.}(2006)Gu, Melnick, Cid~Fernandes, Kunth, Terlevich,
  \& Terlevich}]{gu_emission-line_2006}
Gu Q., Melnick J., Cid~Fernandes R., Kunth D., Terlevich E., Terlevich R.,
  2006, MNRAS, 366, 480

\bibitem[{Hagiwara(2007)}]{hagiwara_low-luminosity_2007}
Hagiwara Y., 2007, AJ, 133, 1176

\bibitem[{Hagiwara {et~al}\mbox{.}(2001)Hagiwara, Henkel, Menten, \&
  Nakai}]{hagiwara_water_2001}
Hagiwara Y., Henkel C., Menten K.~M., Nakai N., 2001, The ApJL, 560, L37

\bibitem[{Ho {et~al}\mbox{.}(1997)Ho, Filippenko, Sargent, \&
  Peng}]{ho_search_1997}
Ho L.~C., Filippenko A.~V., Sargent W. L.~W., Peng C.~Y., 1997, ApJSS, 112, 391

\bibitem[{Ho \& Ulvestad(2001)}]{ho_radio_2001}
Ho L.~C., Ulvestad J.~S., 2001, ApJSS, 133, 77

\bibitem[{H\"onig \& Kishimoto(2010)}]{honig_dusty_2010}
H\"onig S.~F., Kishimoto M., 2010, A\&A, 523, 27

\bibitem[{H\"onig {et~al}\mbox{.}(2012)H\"onig, Kishimoto, Antonucci, Marconi,
  Prieto, Tristram, \& Weigelt}]{honig_parsec-scale_2012}
H\"onig S.~F., Kishimoto M., Antonucci R., Marconi A., Prieto M.~A., Tristram
  K., Weigelt G., 2012, ApJ, 755, 149

\bibitem[{H\"onig {et~al}\mbox{.}(2010)H\"onig, Kishimoto, Gandhi, Smette,
  Asmus, Duschl, Polletta, \& Weigelt}]{honig_dusty_2010-1}
H\"onig S.~F., Kishimoto M., Gandhi P., Smette A., Asmus D., Duschl W.,
  Polletta M., Weigelt G., 2010, A\&A, 515, 23

\bibitem[{H\"onig {et~al}\mbox{.}(2013)H\"onig, Kishimoto, Tristram, Prieto,
  Gandhi, Asmus, Antonucci, Burtscher, Duschl, \& Weigelt}]{honig_dust_2013}
H\"onig S.~F. {et~al.}, 2013, ApJ, 771, 87

\bibitem[{Jackson, Tadhunter \& Sparks(1998)Jackson, Tadhunter, \&
  Sparks}]{jackson_cygnus_1998}
Jackson N., Tadhunter C., Sparks W.~B., 1998, MNRAS, 301, 131

\bibitem[{Jaffe {et~al}\mbox{.}(2004)Jaffe, Meisenheimer, R\"ottgering,
  Leinert, Richichi, Chesneau, Fraix-Burnet, Glazenborg-Kluttig, Granato,
  Graser, Heijligers, K\"ohler, Malbet, Miley, Paresce, Pel, Perrin, Przygodda,
  Schoeller, Sol, Waters, Weigelt, Woillez, \& de~Zeeuw}]{jaffe_central_2004}
Jaffe W. {et~al.}, 2004, Nature, 429, 47

\bibitem[{Jones {et~al}\mbox{.}(1986)Jones, Unwin, Readhead, Sargent,
  Seielstad, Simon, Walker, Benson, Perley, Bridle, Pauliny-Toth, Romney,
  Witzel, Wilkinson, Baath, Booth, Fort, Galt, Mutel, \&
  Linfield}]{jones_high_1986}
Jones D.~L. {et~al.}, 1986, ApJ, 305, 684

\bibitem[{Jorsater, Lindblad \& Boksenberg(1984)Jorsater, Lindblad, \&
  Boksenberg}]{jorsater_kinematics_1984}
Jorsater S., Lindblad P.~O., Boksenberg A., 1984, A\&A, 140, 288

\bibitem[{Kataza {et~al}\mbox{.}(2000)Kataza, Okamoto, Takubo, Onaka, Sako,
  Nakamura, Miyata, \& Yamashita}]{kataza_comics:_2000}
Kataza H., Okamoto Y., Takubo S., Onaka T., Sako S., Nakamura K., Miyata T.,
  Yamashita T., 2000, in {SPIE}, Vol. 4008, Optical and {IR} {Telescope}
  {Instrumentation} and {Detectors}, pp. 1144--1152

\bibitem[{K\"aufl {et~al}\mbox{.}(2015)K\"aufl, Kerber, Asmus, Baksai,
  Di~Lieto, Duhoux, Heikamp, Hummel, Ives, Jakob, Kirchbauer, Mehrgan, Momany,
  Pantin, Pozna, Riquelme, Sandrock, Siebenmorgen, Smette, Stegmeier, Taylor,
  Tristram, Valdes, van~den Ancker, Weilenmann, \& Wolff}]{kaufl_return_2015}
K\"aufl H.~U. {et~al.}, 2015, The Messenger, 159, 15

\bibitem[{Kay(1994)}]{kay_blue_1994}
Kay L.~E., 1994, ApJ, 430, 196

\bibitem[{Kay \& Moran(1998)}]{kay_hidden_1998}
Kay L.~E., Moran E.~C., 1998, PASP, 110, 1003

\bibitem[{Kinney {et~al}\mbox{.}(2000)Kinney, Schmitt, Clarke, Pringle,
  Ulvestad, \& Antonucci}]{kinney_jet_2000}
Kinney A.~L., Schmitt H.~R., Clarke C.~J., Pringle J.~E., Ulvestad J.~S.,
  Antonucci R. R.~J., 2000, ApJ, 537, 152

\bibitem[{Kishimoto {et~al}\mbox{.}(2009)Kishimoto, H\"onig, Antonucci, Kotani,
  Barvainis, Tristram, \& Weigelt}]{kishimoto_exploring_2009}
Kishimoto M., H\"onig S.~F., Antonucci R., Kotani T., Barvainis R., Tristram K.
  R.~W., Weigelt G., 2009, A\&A, 507, L57

\bibitem[{Kishimoto {et~al}\mbox{.}(2011)Kishimoto, H\"onig, Antonucci,
  Millour, Tristram, \& Weigelt}]{kishimoto_mapping_2011}
Kishimoto M., H\"onig S.~F., Antonucci R., Millour F., Tristram K. R.~W.,
  Weigelt G., 2011, A\&A, 536, 78

\bibitem[{Kishimoto {et~al}\mbox{.}(2007)Kishimoto, H\"onig, Beckert, \&
  Weigelt}]{kishimoto_innermost_2007}
Kishimoto M., H\"onig S.~F., Beckert T., Weigelt G., 2007, A\&A, 476, 713

\bibitem[{Kondratko, Greenhill \& Moran(2008)Kondratko, Greenhill, \&
  Moran}]{kondratko_parsec-scale_2008}
Kondratko P.~T., Greenhill L.~J., Moran J.~M., 2008, ApJ, 678, 87

\bibitem[{Kondratko {et~al}\mbox{.}(2006)Kondratko, Greenhill, Moran, Lovell,
  Kuiper, Jauncey, Cameron, G\'omez, Garc\'ia-Mir\'o, Moll,
  de~Gregorio-Monsalvo, \& Jim\'enez-Bail\'on}]{kondratko_discovery_2006}
Kondratko P.~T. {et~al.}, 2006, ApJ, 638, 100

\bibitem[{Krabbe, B\"oker \& Maiolino(2001)Krabbe, B\"oker, \&
  Maiolino}]{krabbe_n-band_2001}
Krabbe A., B\"oker T., Maiolino R., 2001, ApJ, 557, 626

\bibitem[{Kraus {et~al}\mbox{.}(2014)Kraus, Monnier, Harries, Dong, Bate,
  Whitney, Zhu, Buscher, Berger, Haniff, Ireland, Labadie, Lacour, Petrov,
  Ridgway, Surdej, ten Brummelaar, Tuthill, \& van Belle}]{kraus_science_2014}
Kraus S. {et~al.}, 2014, in , eprint: arXiv:1407.7033, p. 914611

\bibitem[{Kristen {et~al}\mbox{.}(1997)Kristen, Jorsater, Lindblad, \&
  Boksenberg}]{kristen_imaging_1997}
Kristen H., Jorsater S., Lindblad P.~O., Boksenberg A., 1997, A\&A, 328, 483

\bibitem[{Kukula {et~al}\mbox{.}(1993)Kukula, Ghosh, Pedlar, Schilizzi, Miley,
  de~Bruyn, \& Saikia}]{kukula_high-resolution_1993}
Kukula M.~J., Ghosh T., Pedlar A., Schilizzi R.~T., Miley G.~K., de~Bruyn
  A.~G., Saikia D.~J., 1993, MNRAS, 264, 893

\bibitem[{Lagage {et~al}\mbox{.}(2004)Lagage, Pel, Authier, Belorgey, Claret,
  Doucet, Dubreuil, Durand, Elswijk, Girardot, K\"aufl, Kroes, Lortholary,
  Lussignol, Marchesi, Pantin, Peletier, Pirard, Pragt, Rio, Schoenmaker,
  Siebenmorgen, Silber, Smette, Sterzik, \& Veyssiere}]{lagage_successful_2004}
Lagage P.~O. {et~al.}, 2004, The Messenger, 117, 12

\bibitem[{Leahy {et~al}\mbox{.}(1997)Leahy, Black, Dennett-Thorpe, Hardcastle,
  Komissarov, Perley, Riley, \& Scheuer}]{leahy_study_1997}
Leahy J.~P., Black A. R.~S., Dennett-Thorpe J., Hardcastle M.~J., Komissarov
  S., Perley R.~A., Riley J.~M., Scheuer P. A.~G., 1997, MNRAS, 291, 20

\bibitem[{Leahy \& Perley(1991)}]{leahy_vla_1991}
Leahy J.~P., Perley R.~A., 1991, AJ, 102, 537

\bibitem[{Leahy \& Perley(1995)}]{leahy_jets_1995}
Leahy J.~P., Perley R.~A., 1995, MNRAS, 277, 1097

\bibitem[{Leipski {et~al}\mbox{.}(2006)Leipski, Falcke, Bennert, \&
  H\"uttemeister}]{leipski_radio_2006}
Leipski C., Falcke H., Bennert N., H\"uttemeister S., 2006, A\&A, 455, 161

\bibitem[{Lipari, Tsvetanov \& Macchetto(1993)Lipari, Tsvetanov, \&
  Macchetto}]{lipari_high-resolution_1993}
Lipari S., Tsvetanov Z., Macchetto F., 1993, ApJ, 405, 186

\bibitem[{Lonsdale {et~al}\mbox{.}(2003)Lonsdale, Lonsdale, Smith, \&
  Diamond}]{lonsdale_vlbi_2003}
Lonsdale C.~J., Lonsdale C.~J., Smith H.~E., Diamond P.~J., 2003, ApJ, 592, 804

\bibitem[{Lopez {et~al}\mbox{.}(2014)Lopez, Lagarde, Jaffe, Petrov, Sch\"oller,
  Antonelli, Beckmann, Berio, Bettonvil, Glindemann, Gonzalez, Graser, Hofmann,
  Millour, Robbe-Dubois, Venema, Wolf, Henning, Lanz, Weigelt, Agocs, Bailet,
  Bresson, Bristow, Dugu\'e, Heininger, Kroes, Laun, Lehmitz, Neumann,
  Augereau, Avila, Behrend, van Belle, Berger, van Boekel, Bonhomme, Bourget,
  Brast, Clausse, Connot, Conzelmann, Cruzalèbes, Csepany, Danchi, Delbo,
  Delplancke, Dominik, van Duin, Elswijk, Fantei, Finger, Gabasch, Gay, Girard,
  Girault, Gitton, Glazenborg, Gont\'e, Guitton, Guniat, De~Haan, Haguenauer,
  Hanenburg, Hogerheijde, ter Horst, Hron, Hugues, Hummel, Idserda, Ives,
  Jakob, Jasko, Jolley, Kiraly, K\"ohler, Kragt, Kroener, Kuindersma, Labadie,
  Leinert, Le~Poole, Lizon, Lucuix, Marcotto, Martinache, Martinot-Lagarde,
  Mathar, Matter, Mauclert, Mehrgan, Meilland, Meisenheimer, Meisner, Mellein,
  Menardi, Menut, Merand, Morel, Mosoni, Navarro, Nussbaum, Ottogalli, Palsa,
  Panduro, Pantin, Parra, Percheron, Duc, Pott, Pozna, Przygodda, Rabbia,
  Richichi, Rigal, Roelfsema, Rupprecht, Schertl, Schmidt, Schuhler, Schuil,
  Spang, Stegmeier, Thiam, Tromp, Vakili, Vannier, Wagner, \&
  Woillez}]{lopez_overview_2014}
Lopez B. {et~al.}, 2014, The Messenger, 157, 5

\bibitem[{Lumsden, Alexander \& Hough(2004)Lumsden, Alexander, \&
  Hough}]{lumsden_spectropolarimetry_2004}
Lumsden S.~L., Alexander D.~M., Hough J.~H., 2004, MNRAS, 348, 1451

\bibitem[{Makarov {et~al}\mbox{.}(2014)Makarov, Prugniel, Terekhova, Courtois,
  \& Vauglin}]{makarov_hyperleda._2014}
Makarov D., Prugniel P., Terekhova N., Courtois H., Vauglin I., 2014, A\&A,
  570, A13

\bibitem[{Marconi {et~al}\mbox{.}(1994)Marconi, Moorwood, Origlia, \&
  Oliva}]{marconi_prominent_1994}
Marconi A., Moorwood A. F.~M., Origlia L., Oliva E., 1994, The Messenger, 78,
  20

\bibitem[{Matthews {et~al}\mbox{.}(1999)Matthews, Gallagher, Krist, Watson,
  Burrows, Griffiths, Hester, Trauger, Ballester, Clarke, Crisp, Evans,
  Hoessel, Holtzman, Mould, Scowen, Stapelfeldt, \&
  Westphal}]{matthews_wfpc2_1999}
Matthews L.~D. {et~al.}, 1999, AJ, 118, 208

\bibitem[{Mayer(1979)}]{mayer_multifrequency_1979}
Mayer C.~J., 1979, MNRAS, 186, 99

\bibitem[{Mediavilla {et~al}\mbox{.}(2005)Mediavilla, Guijarro,
  Castillo-Morales, Jim\'enez-Vicente, Florido, Arribas, Garc\'ia-Lorenzo, \&
  Battaner}]{mediavilla_asymmetrical_2005}
Mediavilla E., Guijarro A., Castillo-Morales A., Jim\'enez-Vicente J., Florido
  E., Arribas S., Garc\'ia-Lorenzo B., Battaner E., 2005, A\&A, 433, 79

\bibitem[{Mel\'endez {et~al}\mbox{.}(2008)Mel\'endez, Kraemer, Armentrout, Deo,
  Crenshaw, Schmitt, Mushotzky, Tueller, Markwardt, \&
  Winter}]{melendez_new_2008}
Mel\'endez M. {et~al.}, 2008, ApJ, 682, 94

\bibitem[{Miller \& Goodrich(1990)}]{miller_spectropolarimetry_1990}
Miller J.~S., Goodrich R.~W., 1990, ApJ, 355, 456

\bibitem[{Momjian {et~al}\mbox{.}(2003)Momjian, Romney, Carilli, \&
  Troland}]{momjian_sensitive_2003}
Momjian E., Romney J.~D., Carilli C.~L., Troland T.~H., 2003, ApJ, 597, 809

\bibitem[{Mor \& Netzer(2012)}]{mor_hot_2012}
Mor R., Netzer H., 2012, MNRAS, 420, 526

\bibitem[{Mor, Netzer \& Elitzur(2009)Mor, Netzer, \& Elitzur}]{mor_dusty_2009}
Mor R., Netzer H., Elitzur M., 2009, ApJ, 705, 298

\bibitem[{Moran {et~al}\mbox{.}(2007)Moran, Barth, Eracleous, \&
  Kay}]{moran_transient_2007}
Moran E.~C., Barth A.~J., Eracleous M., Kay L.~E., 2007, The ApJL, 668, L31

\bibitem[{Moran {et~al}\mbox{.}(2000)Moran, Barth, Kay, \&
  Filippenko}]{moran_frequency_2000}
Moran E.~C., Barth A.~J., Kay L.~E., Filippenko A.~V., 2000, The ApJL, 540, L73

\bibitem[{Morganti, Oosterloo \& Tsvetanov(1998)Morganti, Oosterloo, \&
  Tsvetanov}]{morganti_radio_1998}
Morganti R., Oosterloo T., Tsvetanov Z., 1998, AJ, 115, 915

\bibitem[{Morganti {et~al}\mbox{.}(1991)Morganti, Robinson, Fosbury,
  di~Serego~Alighieri, Tadhunter, \& Malin}]{morganti_nature_1991}
Morganti R., Robinson A., Fosbury R. A.~E., di~Serego~Alighieri S., Tadhunter
  C.~N., Malin D.~F., 1991, MNRAS, 249, 91

\bibitem[{Morganti {et~al}\mbox{.}(1999)Morganti, Tsvetanov, Gallimore, \&
  Allen}]{morganti_radio_1999}
Morganti R., Tsvetanov Z.~I., Gallimore J., Allen M.~G., 1999, A\&A Supplement
  Series, 137, 457

\bibitem[{Morris {et~al}\mbox{.}(1985)Morris, Ward, Whittle, Wilson, \&
  Taylor}]{morris_velocity_1985}
Morris S., Ward M., Whittle M., Wilson A.~S., Taylor K., 1985, MNRAS, 216, 193

\bibitem[{Mulchaey {et~al}\mbox{.}(1994)Mulchaey, Wilson, Bower, Heckman,
  Krolik, \& Miley}]{mulchaey_hubble_1994}
Mulchaey J.~S., Wilson A.~S., Bower G.~A., Heckman T.~M., Krolik J.~H., Miley
  G.~K., 1994, ApJ, 433, 625

\bibitem[{Mulchaey, Wilson \& Tsvetanov(1996)Mulchaey, Wilson, \&
  Tsvetanov}]{mulchaey_emission-line_1996}
Mulchaey J.~S., Wilson A.~S., Tsvetanov Z., 1996, ApJSS, 102, 309

\bibitem[{M\"uller-S\'anchez {et~al}\mbox{.}(2011)M\"uller-S\'anchez, Prieto,
  Hicks, Vives-Arias, Davies, Malkan, Tacconi, \&
  Genzel}]{muller-sanchez_outflows_2011}
M\"uller-S\'anchez F., Prieto M.~A., Hicks E. K.~S., Vives-Arias H., Davies
  R.~I., Malkan M., Tacconi L.~J., Genzel R., 2011, ApJ, 739, 69

\bibitem[{Mundell {et~al}\mbox{.}(1995)Mundell, Holloway, Pedlar, Meaburn,
  Kukula, \& Axon}]{mundell_anisotropic_1995}
Mundell C.~G., Holloway A.~J., Pedlar A., Meaburn J., Kukula M.~J., Axon D.~J.,
  1995, MNRAS, 275, 67

\bibitem[{Mundell {et~al}\mbox{.}(2000)Mundell, Wilson, Ulvestad, \&
  Roy}]{mundell_parsec-scale_2000}
Mundell C.~G., Wilson A.~S., Ulvestad J.~S., Roy A.~L., 2000, ApJ, 529, 816

\bibitem[{Nagao {et~al}\mbox{.}(2004)Nagao, Kawabata, Murayama, Ohyama,
  Taniguchi, Sumiya, \& Sasaki}]{nagao_detection_2004}
Nagao T., Kawabata K.~S., Murayama T., Ohyama Y., Taniguchi Y., Sumiya R.,
  Sasaki S.~S., 2004, AJ, 128, 109

\bibitem[{Nagar {et~al}\mbox{.}(1999)Nagar, Wilson, Mulchaey, \&
  Gallimore}]{nagar_radio_1999}
Nagar N.~M., Wilson A.~S., Mulchaey J.~S., Gallimore J.~F., 1999, ApJSS, 120,
  209

\bibitem[{Nenkova {et~al}\mbox{.}(2008)Nenkova, Sirocky, Ivezi\'c, \&
  Elitzur}]{nenkova_agn_2008}
Nenkova M., Sirocky M.~M., Ivezi\'c v., Elitzur M., 2008, ApJ, 685, 147

\bibitem[{Netzer(2015)}]{netzer_revisiting_2015}
Netzer H., 2015, arXiv:1505.00811 [astro-ph], arXiv: 1505.00811

\bibitem[{Netzer \& Laor(1993)}]{netzer_dust_1993}
Netzer H., Laor A., 1993, The ApJL, 404, L51

\bibitem[{Ogle, Whysong \& Antonucci(2006)Ogle, Whysong, \&
  Antonucci}]{ogle_spitzer_2006}
Ogle P., Whysong D., Antonucci R., 2006, ApJ, 647, 161

\bibitem[{Oliva {et~al}\mbox{.}(1998)Oliva, Marconi, Cimatti, \&
  Alighieri}]{oliva_spectropolarimetry_1998}
Oliva E., Marconi A., Cimatti A., Alighieri S. D.~S., 1998, A\&A, 329, L21

\bibitem[{Packham {et~al}\mbox{.}(2005)Packham, Radomski, Roche, Aitken,
  Perlman, Alonso-Herrero, Colina, \& Telesco}]{packham_extended_2005}
Packham C., Radomski J.~T., Roche P.~F., Aitken D.~K., Perlman E.,
  Alonso-Herrero A., Colina L., Telesco C.~M., 2005, The ApJL, 618, L17

\bibitem[{Pedlar {et~al}\mbox{.}(1990)Pedlar, Ghataure, Davies, Harrison,
  Perley, Crane, \& Unger}]{pedlar_radio_1990}
Pedlar A., Ghataure H.~S., Davies R.~D., Harrison B.~A., Perley R., Crane
  P.~C., Unger S.~W., 1990, MNRAS, 246, 477

\bibitem[{Pereira-Santaella {et~al}\mbox{.}(2010)Pereira-Santaella,
  Diamond-Stanic, Alonso-Herrero, \&
  Rieke}]{pereira-santaella_mid-infrared_2010}
Pereira-Santaella M., Diamond-Stanic A.~M., Alonso-Herrero A., Rieke G.~H.,
  2010, ApJ, 725, 2270

\bibitem[{Perez {et~al}\mbox{.}(1989)Perez, Gonzalez-Delgado, Tadhunter, \&
  Tsvetanov}]{perez_complex_1989}
Perez E., Gonzalez-Delgado R., Tadhunter C., Tsvetanov Z., 1989, MNRAS, 241,
  31P

\bibitem[{Perley, Bridle \& Willis(1984)Perley, Bridle, \&
  Willis}]{perley_high-resolution_1984}
Perley R.~A., Bridle A.~H., Willis A.~G., 1984, ApJSS, 54, 291

\bibitem[{Perley, Dreher \& Cowan(1984)Perley, Dreher, \&
  Cowan}]{perley_jet_1984}
Perley R.~A., Dreher J.~W., Cowan J.~J., 1984, The ApJL, 285, L35

\bibitem[{Perley, Roser \& Meisenheimer(1997)Perley, Roser, \&
  Meisenheimer}]{perley_radio_1997}
Perley R.~A., Roser H.-J., Meisenheimer K., 1997, A\&A, 328, 12

\bibitem[{Phillips {et~al}\mbox{.}(1983)Phillips, Edmunds, Pagel, \&
  Turtle}]{phillips_remarkable_1983}
Phillips M.~M., Edmunds M.~G., Pagel B. E.~J., Turtle A.~J., 1983, MNRAS, 203,
  759

\bibitem[{Pogge(1988)}]{pogge_extended_1988}
Pogge R.~W., 1988, ApJ, 328, 519

\bibitem[{Pogge(1989)}]{pogge_circumnuclear_1989}
Pogge R.~W., 1989, ApJ, 345, 730

\bibitem[{Pogge \& De~Robertis(1995)}]{pogge_imaging_1995}
Pogge R.~W., De~Robertis M.~M., 1995, ApJ, 451, 585

\bibitem[{Privon {et~al}\mbox{.}(2012)Privon, Baum, O'Dea, Gallimore,
  Noel-Storr, Axon, \& Robinson}]{privon_modeling_2012}
Privon G.~C., Baum S.~A., O'Dea C.~P., Gallimore J., Noel-Storr J., Axon D.~J.,
  Robinson A., 2012, ApJ, 747, 46

\bibitem[{Raban {et~al}\mbox{.}(2009)Raban, Jaffe, R\"ottgering, Meisenheimer,
  \& Tristram}]{raban_resolving_2009}
Raban D., Jaffe W., R\"ottgering H., Meisenheimer K., Tristram K. R.~W., 2009,
  MNRAS, 394, 1325

\bibitem[{Radomski {et~al}\mbox{.}(2003)Radomski, Piña, Packham, Telesco,
  De~Buizer, Fisher, \& Robinson}]{radomski_resolved_2003}
Radomski J.~T., Piña R.~K., Packham C., Telesco C.~M., De~Buizer J.~M., Fisher
  R.~S., Robinson A., 2003, ApJ, 587, 117

\bibitem[{Radomski {et~al}\mbox{.}(2002)Radomski, Piña, Packham, Telesco, \&
  Tadhunter}]{radomski_high-resolution_2002}
Radomski J.~T., Piña R.~K., Packham C., Telesco C.~M., Tadhunter C.~N., 2002,
  ApJ, 566, 675

\bibitem[{Reunanen, Prieto \& Siebenmorgen(2010)Reunanen, Prieto, \&
  Siebenmorgen}]{reunanen_vlt_2010}
Reunanen J., Prieto M.~A., Siebenmorgen R., 2010, MNRAS, 402, 879

\bibitem[{Sandqvist, Joersaeter \& Lindblad(1995)Sandqvist, Joersaeter, \&
  Lindblad}]{sandqvist_central_1995}
Sandqvist A., Joersaeter S., Lindblad P.~O., 1995, A\&A, 295, 585

\bibitem[{Schmid, Appenzeller \& Burch(2003)Schmid, Appenzeller, \&
  Burch}]{schmid_spectropolarimetry_2003}
Schmid H.~M., Appenzeller I., Burch U., 2003, A\&A, 404, 505

\bibitem[{Schmitt {et~al}\mbox{.}(2003{\natexlab{a}})Schmitt, Donley,
  Antonucci, Hutchings, \& Kinney}]{schmitt_hubble_2003}
Schmitt H.~R., Donley J.~L., Antonucci R. R.~J., Hutchings J.~B., Kinney A.~L.,
  2003{\natexlab{a}}, ApJSS, 148, 327

\bibitem[{Schmitt {et~al}\mbox{.}(2003{\natexlab{b}})Schmitt, Donley,
  Antonucci, Hutchings, Kinney, \& Pringle}]{schmitt_hubble_2003-1}
Schmitt H.~R., Donley J.~L., Antonucci R. R.~J., Hutchings J.~B., Kinney A.~L.,
  Pringle J.~E., 2003{\natexlab{b}}, ApJ, 597, 768

\bibitem[{Schmitt \& Kinney(1996)}]{schmitt_comparison_1996}
Schmitt H.~R., Kinney A.~L., 1996, ApJ, 463, 498

\bibitem[{Schmitt \& Storchi-Bergmann(1995)}]{schmitt_anisotropic_1995}
Schmitt H.~R., Storchi-Bergmann T., 1995, MNRAS, 276, 592

\bibitem[{Schmitt {et~al}\mbox{.}(2001)Schmitt, Ulvestad, Antonucci, \&
  Kinney}]{schmitt_jet_2001}
Schmitt H.~R., Ulvestad J.~S., Antonucci R. R.~J., Kinney A.~L., 2001, ApJSS,
  132, 199

\bibitem[{Schommer {et~al}\mbox{.}(1988)Schommer, Caldwell, Wilson, Baldwin,
  Phillips, Williams, \& Turtle}]{schommer_ionized_1988}
Schommer R.~A., Caldwell N., Wilson A.~S., Baldwin J.~A., Phillips M.~M.,
  Williams T.~B., Turtle A.~J., 1988, ApJ, 324, 154

\bibitem[{Schreier, Burns \& Feigelson(1981)Schreier, Burns, \&
  Feigelson}]{schreier_detection_1981}
Schreier E.~J., Burns J.~O., Feigelson E.~D., 1981, ApJ, 251, 523

\bibitem[{Schweitzer {et~al}\mbox{.}(2006)Schweitzer, Lutz, Sturm, Contursi,
  Tacconi, Lehnert, Dasyra, Genzel, Veilleux, Rupke, Kim, Baker, Netzer,
  Sternberg, Mazzarella, \& Lord}]{schweitzer_spitzer_2006}
Schweitzer M. {et~al.}, 2006, ApJ, 649, 79

\bibitem[{Simkin {et~al}\mbox{.}(1999)Simkin, Sadler, Sault, Tingay, \&
  Callcut}]{simkin_pictor_1999}
Simkin S.~M., Sadler E.~M., Sault R., Tingay S.~J., Callcut J., 1999, ApJSS,
  123, 447

\bibitem[{Simpson {et~al}\mbox{.}(1997)Simpson, Wilson, Bower, Heckman, Krolik,
  \& Miley}]{simpson_one-sided_1997}
Simpson C., Wilson A.~S., Bower G., Heckman T.~M., Krolik J.~H., Miley G.~K.,
  1997, ApJ, 474, 121

\bibitem[{Smaji\'c {et~al}\mbox{.}(2012)Smaji\'c, Fischer, Zuther, \&
  Eckart}]{smajic_unveiling_2012}
Smaji\'c S., Fischer S., Zuther J., Eckart A., 2012, A\&A, 544, 105

\bibitem[{Smith {et~al}\mbox{.}(2004)Smith, Robinson, Alexander, Young, Axon,
  \& Corbett}]{smith_seyferts_2004}
Smith J.~E., Robinson A., Alexander D.~M., Young S., Axon D.~J., Corbett E.~A.,
  2004, MNRAS, 350, 140

\bibitem[{Stalevski {et~al}\mbox{.}(2012)Stalevski, Fritz, Baes, Nakos, \&
  Popovi\'c}]{stalevski_3D_2012}
Stalevski M., Fritz J., Baes M., Nakos T., Popovi\'c L.~Ä., 2012, MNRAS, 420,
  2756

\bibitem[{Stern, Laor \& Baskin(2014)Stern, Laor, \&
  Baskin}]{stern_radiation_2014}
Stern J., Laor A., Baskin A., 2014, MNRAS, 438, 901

\bibitem[{Storchi-Bergmann \& Bonatto(1991)}]{storchi-bergmann_detection_1991}
Storchi-Bergmann T., Bonatto C.~J., 1991, MNRAS, 250, 138

\bibitem[{Tadhunter \& Tsvetanov(1989)}]{tadhunter_anisotropic_1989}
Tadhunter C., Tsvetanov Z., 1989, Nature, 341, 422

\bibitem[{Telesco {et~al}\mbox{.}(1998)Telesco, Pina, Hanna, Julian, Hon, \&
  Kisko}]{telesco_gatircam:_1998}
Telesco C.~M., Pina R.~K., Hanna K.~T., Julian J.~A., Hon D.~B., Kisko T.~M.,
  1998, in {SPIE}, Vol. 3354, Infrared {Astronomical} {Instrumentation}, pp.
  534--544

\bibitem[{Tommasin {et~al}\mbox{.}(2010)Tommasin, Spinoglio, Malkan, \&
  Fazio}]{tommasin_spitzer-irs_2010}
Tommasin S., Spinoglio L., Malkan M.~A., Fazio G., 2010, ApJ, 709, 1257

\bibitem[{Tran(1995)}]{tran_nature_1995}
Tran H.~D., 1995, ApJ, 440, 578

\bibitem[{Tristram {et~al}\mbox{.}(2014)Tristram, Burtscher, Jaffe,
  Meisenheimer, H\"onig, Kishimoto, Schartmann, \&
  Weigelt}]{tristram_dusty_2014}
Tristram K. R.~W., Burtscher L., Jaffe W., Meisenheimer K., H\"onig S.~F.,
  Kishimoto M., Schartmann M., Weigelt G., 2014, A\&A, 563, 82

\bibitem[{Tristram {et~al}\mbox{.}(2007)Tristram, Meisenheimer, Jaffe,
  Schartmann, Rix, Leinert, Morel, Wittkowski, R\"ottgering, Perrin, Lopez,
  Raban, Cotton, Graser, Paresce, \& Henning}]{tristram_resolving_2007}
Tristram K. R.~W. {et~al.}, 2007, A\&A, 474, 837

\bibitem[{Ueda {et~al}\mbox{.}(2015)Ueda, Hashimoto, Ichikawa, Ishino, Kniazev,
  V\"ais\"anen, Ricci, Berney, Gandhi, Koss, Mushotzky, Terashima,
  Trakhtenbrot, \& Crenshaw}]{ueda_[o_2015}
Ueda Y. {et~al.}, 2015, ApJ, 815, 1

\bibitem[{Ulvestad \& Wilson(1983)}]{ulvestad_nuclear_1983}
Ulvestad J.~S., Wilson A.~S., 1983, The ApJL, 264, L7

\bibitem[{Ulvestad \& Wilson(1984{\natexlab{a}})}]{ulvestad_radio_1984-1}
Ulvestad J.~S., Wilson A.~S., 1984{\natexlab{a}}, ApJ, 278, 544

\bibitem[{Ulvestad \& Wilson(1984{\natexlab{b}})}]{ulvestad_radio_1984}
Ulvestad J.~S., Wilson A.~S., 1984{\natexlab{b}}, ApJ, 285, 439

\bibitem[{Ulvestad \& Wilson(1989)}]{ulvestad_radio_1989}
Ulvestad J.~S., Wilson A.~S., 1989, ApJ, 343, 659

\bibitem[{Vermeulen, Readhead \& Backer(1994)Vermeulen, Readhead, \&
  Backer}]{vermeulen_discovery_1994}
Vermeulen R.~C., Readhead A. C.~S., Backer D.~C., 1994, The ApJL, 430, L41

\bibitem[{Walker, Benson \& Unwin(1987)Walker, Benson, \&
  Unwin}]{walker_radio_1987}
Walker R.~C., Benson J.~M., Unwin S.~C., 1987, ApJ, 316, 546

\bibitem[{Walker, Romney \& Benson(1994)Walker, Romney, \&
  Benson}]{walker_detection_1994}
Walker R.~C., Romney J.~D., Benson J.~M., 1994, The ApJL, 430, L45

\bibitem[{Ward {et~al}\mbox{.}(1980)Ward, Penston, Blades, \&
  Turtle}]{ward_new_1980}
Ward M., Penston M.~V., Blades J.~C., Turtle A.~J., 1980, MNRAS, 193, 563

\bibitem[{Weaver {et~al}\mbox{.}(2010)Weaver, Mel\'endez, Mushotzky, Kraemer,
  Engle, Malumuth, Tueller, Markwardt, Berghea, Dudik, Winter, \&
  Armus}]{weaver_mid-infrared_2010}
Weaver K.~A. {et~al.}, 2010, ApJ, 716, 1151

\bibitem[{Weedman {et~al}\mbox{.}(2012)Weedman, Sargsyan, Lebouteiller, Houck,
  \& Barry}]{weedman_infrared_2012}
Weedman D., Sargsyan L., Lebouteiller V., Houck J., Barry D., 2012, ApJ, 761,
  184

\bibitem[{Wehrle \& Morris(1988)}]{wehrle_radio_1988}
Wehrle A.~E., Morris M., 1988, AJ, 95, 1689

\bibitem[{Whysong \& Antonucci(2004)}]{whysong_thermal_2004}
Whysong D., Antonucci R., 2004, ApJ, 602, 116

\bibitem[{Willett {et~al}\mbox{.}(2011)Willett, Darling, Spoon, Charmandaris,
  \& Armus}]{willett_mid-infrared_2011}
Willett K.~W., Darling J., Spoon H. W.~W., Charmandaris V., Armus L., 2011,
  ApJSS, 193, 18

\bibitem[{Wilson {et~al}\mbox{.}(1993)Wilson, Braatz, Heckman, Krolik, \&
  Miley}]{wilson_ionization_1993}
Wilson A.~S., Braatz J.~A., Heckman T.~M., Krolik J.~H., Miley G.~K., 1993, The
  ApJL, 419, L61

\bibitem[{Wilson {et~al}\mbox{.}(2000)Wilson, Shopbell, Simpson,
  Storchi-Bergmann, Barbosa, \& Ward}]{wilson_hubble_2000}
Wilson A.~S., Shopbell P.~L., Simpson C., Storchi-Bergmann T., Barbosa F.
  K.~B., Ward M.~J., 2000, AJ, 120, 1325

\bibitem[{Wilson \& Tsvetanov(1994)}]{wilson_ionization_1994}
Wilson A.~S., Tsvetanov Z.~I., 1994, AJ, 107, 1227

\bibitem[{Wilson \& Ulvestad(1982{\natexlab{a}})}]{wilson_radio_1982}
Wilson A.~S., Ulvestad J.~S., 1982{\natexlab{a}}, ApJ, 260, 56

\bibitem[{Wilson \& Ulvestad(1982{\natexlab{b}})}]{wilson_radio_1982-1}
Wilson A.~S., Ulvestad J.~S., 1982{\natexlab{b}}, ApJ, 263, 576

\bibitem[{Wilson {et~al}\mbox{.}(1989)Wilson, Wu, Heckman, Baldwin, \&
  Balick}]{wilson_kinematics_1989}
Wilson A.~S., Wu X., Heckman T.~M., Baldwin J.~A., Balick B., 1989, ApJ, 339,
  729

\bibitem[{Wu, Zhao \& Meng(2011)Wu, Zhao, \& Meng}]{wu_diagnostics_2011}
Wu Y.-Z., Zhao Y.-H., Meng X.-M., 2011, 1104.5707

\bibitem[{Yankulova, Golev \& Jockers(2007)Yankulova, Golev, \&
  Jockers}]{yankulova_luminous_2007}
Yankulova I.~M., Golev V.~K., Jockers K., 2007, A\&A, 469, 891

\bibitem[{Young {et~al}\mbox{.}(1996{\natexlab{a}})Young, Hough, Efstathiou,
  Wills, Axon, Bailey, \& Ward}]{young_scattered_1996}
Young S., Hough J.~H., Efstathiou A., Wills B.~J., Axon D.~J., Bailey J.~A.,
  Ward M.~J., 1996{\natexlab{a}}, MNRAS, 279, L72

\bibitem[{Young {et~al}\mbox{.}(1996{\natexlab{b}})Young, Hough, Efstathiou,
  Wills, Bailey, Ward, \& Axon}]{young_polarimetry_1996}
Young S., Hough J.~H., Efstathiou A., Wills B.~J., Bailey J.~A., Ward M.~J.,
  Axon D.~J., 1996{\natexlab{b}}, MNRAS, 281, 1206

\bibitem[{Zirbel \& Baum(1998)}]{zirbel_ultraviolet_1998}
Zirbel E.~L., Baum S.~A., 1998, ApJSS, 114, 177

\end{thebibliography}

\appendix
\newpage
\clearpage

{%\onecolumn

\LongTables

\begin{landscape}
\tabletypesize{\tiny}
\begin{deluxetable}{l		c		c		c		c		c		c		c		c		c		c		c		c		c		c		c		c		c	 c  c c}
%\tiny

\tablewidth{0pt}
\tablecaption{Properties of the Seyfert sample.\label{tab:sam}} % title of Table
\tablehead{
\colhead{		} & \colhead{		} & \colhead{	Opt.	} & \colhead{	nuc.	} & \colhead{		ext.	} & \colhead{	MIR			} & \colhead{	NLR	} & \colhead{	NLR	} & \colhead{	radio	} & \colhead{	radio	} & \colhead{	pol.	} & \colhead{	pol.	} & \colhead{	mas.	} & \colhead{	mas.	} & \colhead{	SA			} & \colhead{	host	} & \colhead{	host	} & \colhead{		$\log F$	} & \colhead{	\oiv	}\\
\colhead{	Object	} & \colhead{	$D$	} & \colhead{	class	} & \colhead{	MIR	} & \colhead{		nuc.	} & \colhead{	PA			} & \colhead{	PA	} & \colhead{	ref.	} & \colhead{	PA	} & \colhead{	ref.	} & \colhead{	PA	} & \colhead{	ref.	} & \colhead{	PA	} & \colhead{	ref.	} & \colhead{	PA			} & \colhead{	PA	} & \colhead{	ref.	} & \colhead{		(\oiv)	} & \colhead{	ref.	}\\
\colhead{		} & \colhead{	[Mpc]	} & \colhead{		} & \colhead{	ext.	} & \colhead{		ratio	} & \colhead{	[deg]			} & \colhead{	[deg]	} & \colhead{		} & \colhead{	[deg]	} & \colhead{		} & \colhead{	[deg]	} & \colhead{		} & \colhead{	[deg]	} & \colhead{		} & \colhead{	[deg]			} & \colhead{	[deg]	} & \colhead{		} & \colhead{		 [erg\,s$^{-1}$\,cm$^{-2}$]	} & \colhead{		}\\
\colhead{	(1)	} & \colhead{	(2)	} & \colhead{	(3)	} & \colhead{	(4)	} & \colhead{		(5)	} & \colhead{	(6)			} & \colhead{	(7)	} & \colhead{	(8)	} & \colhead{	(9)	} & \colhead{	(10)	} & \colhead{	(11)	} & \colhead{	(12)	} & \colhead{	(13)	} & \colhead{	(14)	} & \colhead{	(15)			} & \colhead{	(16)	} & \colhead{	(17)	} & \colhead{		(18)	} & \colhead{	(19)	}\\
}
\startdata	
				
%newname&D&C&morph.&totnuc&MIRPA&NLRPA&NLRref&RADPA&RADref&POLPA&POLref&MASPA&MASref&SAPA&HPA&Href&FOIV&OIVref	\\
1H 0419-577&499.0&1.5&n&0.16&44	$\pm$	48&\nodata&1&\nodata&1&\nodata&1&\nodata&1&\nodata&\nodata&1&\nodata&1\\
1RXS J112716.6+&512.0&1.8&u&0.99&167	$\pm$	13&\nodata&1&\nodata&1&\nodata&1&\nodata&1&\nodata&\nodata&1&\nodata&1\\
2MASX J03565655&351.0&1.9&u&1.60&113	$\pm$	19&\nodata&1&\nodata&1&\nodata&1&\nodata&1&\nodata&72&2&\nodata&1\\
2MASX J09180027&781.0&2&u&0.24&97&\nodata&1&\nodata&1&\nodata&1&\nodata&1&\nodata&\nodata&1&\nodata&1\\
3C 33&273.0&2&p&1.28&121&\nodata&1&200&3, 4&150&5&\nodata&1&40&\nodata&1&-13.09&6\\
3C 98&137.0&2&u&\nodata&74&\nodata&1&20&7, 8&\nodata&1&\nodata&1&20&54&2&-13.35&9\\
3C 105&421.0&2&u&\nodata&146	$\pm$	63&\nodata&1&200&7, 8&\nodata&1&\nodata&1&200&\nodata&1&-13.73&9\\
3C 120&150.0&1.5&u&0.18&143	$\pm$	35&\nodata&1&85&10&\nodata&1&\nodata&1&85&119&2&-11.95&11\\
3C 227&414.0&1.5&n&0.01&149&\nodata&1&90&12&\nodata&1&\nodata&1&90&188&2&-13.65&9\\
3C 273&792.0&1&n&0.01&114	$\pm$	34&\nodata&1&\nodata&1&\nodata&1&\nodata&1&\nodata&\nodata&1&-13.07&11\\
3C 321&460.0&2&p&2.59&89&\nodata&1&135&7&50&13&\nodata&1&138	$\pm$	4&6&2&-12.13&9\\
3C 327&505.0&1&p&0.45&97&\nodata&1&100&7, 8&\nodata&1&\nodata&1&100&137&2&-12.42&9\\
3C 382&267.0&1&n&\nodata&150&\nodata&1&230&12&\nodata&1&\nodata&1&230&90&2&-13.52&9\\
3C 390.3&259.0&1.5&p&0.75&92&\nodata&1&145&14&\nodata&1&\nodata&1&145&90&2&-13.59&9\\
3C 403&271.0&2&u&\nodata&60&\nodata&1&70&12&\nodata&1&\nodata&1&70&-9&2&-12.80&9\\
3C 445&254.0&1.5&u&0.23&85	$\pm$	39&\nodata&1&80&8&\nodata&1&\nodata&1&80&73&2&-12.65&15\\
3C 452&378.0&2&u&0.51&141&\nodata&1&75&12&\nodata&1&\nodata&1&75&99&2&-13.89&6\\
4C +73.08&271.0&2&u&0.53&169&\nodata&1&245&16&\nodata&1&\nodata&1&245&199&2&\nodata&1\\
Ark 120&149.0&1&n&0.05&66	$\pm$	46&\nodata&1&\nodata&1&\nodata&1&\nodata&1&\nodata&-16&2&-13.43&11\\
\rowstyle{\bfseries} Cen A&3.8&2&y&\nodata&74	$\pm$	42&55&17&55&18&\nodata&1&\nodata&1&55&33&2&-12.00&19\\
\rowstyle{\bfseries} Circinus&4.2&2&y&0.40&100	$\pm$	10&135&20, 21&\nodata&1&68&22&56&23&146	$\pm$	12&36&2&-11.17&19\\
\rowstyle{\bfseries} Cygnus A&257.0&2&y&0.68&94	$\pm$	10&105&24, 25&105&26&\nodata&1&\nodata&1&105&151&2&-12.34&27\\
ESO 5-4&22.4&2&u&0.29&82	$\pm$	25&\nodata&1&\nodata&1&\nodata&1&\nodata&1&\nodata&93&2&-13.35&28\\
ESO 33-2&82.3&2&p&0.61&133&185&29&\nodata&1&\nodata&1&\nodata&1&185&175&2&-12.86&28\\
ESO 103-35&59.5&2&p&1.24&109	$\pm$	12&\nodata&1&\nodata&1&\nodata&1&\nodata&1&\nodata&42&2&-12.51&11\\
ESO 121-28&187.0&2&n&\nodata&25	$\pm$	9&\nodata&1&\nodata&1&\nodata&1&\nodata&1&\nodata&89&30&\nodata&1\\
ESO 138-1&41.8&2&u&0.25&121	$\pm$	21&140&31&215&32&\nodata&1&\nodata&1&178	$\pm$	53&141&2&\nodata&1\\
ESO 141-55&169.0&1.2&p&0.24&120	$\pm$	47&\nodata&1&\nodata&1&\nodata&1&\nodata&1&\nodata&71&2&-13.14&33\\
ESO 198-24&208.0&1&n&\nodata&98	$\pm$	89&\nodata&1&\nodata&1&\nodata&1&\nodata&1&\nodata&\nodata&1&\nodata&1\\
ESO 209-12&189.0&1.5&u&0.13&132	$\pm$	57&\nodata&1&\nodata&1&\nodata&1&\nodata&1&\nodata&65&2&\nodata&1\\
ESO 253-3&196.0&2&u&0.06&125&\nodata&1&\nodata&1&\nodata&1&\nodata&1&\nodata&162&2&-12.61&15\\
ESO 263-13&158.0&2&n&0.13&96	$\pm$	24&\nodata&1&\nodata&1&\nodata&1&\nodata&1&\nodata&22&2&\nodata&1\\
ESO 297-18&111.0&2&u&\nodata&149	$\pm$	50&\nodata&1&\nodata&1&\nodata&1&\nodata&1&\nodata&144&2&-13.48&34\\
ESO 323-32&76.4&1.9&u&\nodata&76	$\pm$	33&\nodata&1&\nodata&1&\nodata&1&\nodata&1&\nodata&126&2&\nodata&1\\
\rowstyle{\bfseries} ESO 323-77&71.8&1.2&y&0.10&95	$\pm$	27&180&35&\nodata&1&84&36&\nodata&1&180&156&37&-12.60&28\\
ESO 362-18&56.5&1.5&n&0.01&123	$\pm$	45&135&38&\nodata&1&\nodata&1&\nodata&1&135&145&2&-12.92&33\\
ESO 416-2&272.0&1.9&u&\nodata&44	$\pm$	21&\nodata&1&\nodata&1&\nodata&1&\nodata&1&\nodata&-13&2&\nodata&1\\
\rowstyle{\bfseries} ESO 428-14&28.2&2&y&0.85&135	$\pm$	20&130&39&\nodata&1&\nodata&1&\nodata&1&130&137&2&-11.51&30\\
ESO 506-27&119.0&2&n&0.05&93	$\pm$	8&\nodata&1&\nodata&1&\nodata&1&\nodata&1&\nodata&69&2&-13.41&34\\
ESO 511-30&105.0&1&p&\nodata&82	$\pm$	3&\nodata&1&\nodata&1&\nodata&1&\nodata&1&\nodata&\nodata&1&-14.10&34\\
ESO 548-81&63.5&1&p&0.67&72	$\pm$	46&\nodata&1&\nodata&1&\nodata&1&\nodata&1&\nodata&141&2&-13.59&34\\
Fairall 9&215.0&1.2&n&0.04&94	$\pm$	62&\nodata&1&\nodata&1&\nodata&1&\nodata&1&\nodata&115&2&-13.20&11\\
Fairall 49&90.1&2&p&1.47&42	$\pm$	19&\nodata&1&\nodata&1&15&40&\nodata&1&\nodata&120&2&-12.41&11\\
Fairall 51&64.1&1.5&u&\nodata&180&160&29&\nodata&1&\nodata&1&\nodata&1&160&169&2&-12.61&11\\
H 0557-385&156.0&1.2&u&\nodata&115	$\pm$	38&\nodata&1&\nodata&1&\nodata&1&\nodata&1&\nodata&132&2&-13.27&15\\
H1143-182&156.0&1.5&n&0.01&78	$\pm$	9&\nodata&1&\nodata&1&\nodata&1&\nodata&1&\nodata&40&2&-12.62&34\\
IC 3639&53.6&2&n&0.09&105	$\pm$	32&\nodata&1&\nodata&1&\nodata&1&\nodata&1&\nodata&33&30&$\le$	-12.45&19\\
IC 4329A&76.5&1.2&p&\nodata&65	$\pm$	9&36&35&\nodata&1&\nodata&1&\nodata&1&36&46&2&-11.93&15\\
IC 4518W&76.1&2&p&0.10&158&\nodata&1&\nodata&1&\nodata&1&\nodata&1&\nodata&135&2&-12.07&41\\
\rowstyle{\bfseries} IC 5063&49.1&2&y&0.19&108	$\pm$	5&115&29&115&42&8&40&\nodata&1&107	$\pm$	12&120&2&-11.94&15\\
IRAS 01003-2238&565.0&2&u&\nodata&81&\nodata&1&\nodata&1&\nodata&1&\nodata&1&\nodata&118&2&$\le$	-14.52&43\\
IRAS 04103-2838&567.0&2&p&0.13&77&\nodata&1&\nodata&1&\nodata&1&\nodata&1&\nodata&50&2&-13.37&33\\
IRAS 05189-2524&196.0&2&p&0.74&94	$\pm$	4&\nodata&1&\nodata&1&50&44&\nodata&1&140&107&2&-12.58&45\\
IRAS 09149-6206&269.0&1&n&\nodata&80	$\pm$	12&\nodata&1&\nodata&1&\nodata&1&\nodata&1&\nodata&\nodata&1&\nodata&1\\
IRAS 13349+2438&522.0&1n&u&\nodata&16	$\pm$	19&\nodata&1&\nodata&1&\nodata&1&\nodata&1&\nodata&\nodata&1&-13.14&46\\
I Zw 1&269.0&1n&u&0.31&76	$\pm$	66&\nodata&1&\nodata&1&\nodata&1&\nodata&1&\nodata&156&2&-13.57&33\\
LEDA 170194&173.0&2&n&0.04&44	$\pm$	18&\nodata&1&\nodata&1&\nodata&1&\nodata&1&\nodata&8&2&\nodata&1\\
LEDA 178130&160.0&2&p&0.63&118	$\pm$	39&\nodata&1&\nodata&1&\nodata&1&\nodata&1&\nodata&72&2&-13.74&34\\
LEDA 549777&284.1&2&u&\nodata&123&\nodata&1&\nodata&1&\nodata&1&\nodata&1&\nodata&206&2&\nodata&1\\
M51a&8.1&2&p&1.33&52	$\pm$	10&-17&47&-20&48&\nodata&1&155&49, 50&-19	$\pm$	2&-17&2&-12.61&19\\
MCG-1-5-47&73.4&2&u&\nodata&35	$\pm$	59&\nodata&1&\nodata&1&\nodata&1&\nodata&1&\nodata&-20&2&\nodata&1\\
MCG-1-13-25&71.2&1.2&p&0.55&56	$\pm$	45&\nodata&1&\nodata&1&\nodata&1&\nodata&1&\nodata&35&2&-14.12&28\\
MCG-1-24-12&93.8&2&n&0.23&122	$\pm$	46&75&29&\nodata&1&\nodata&1&\nodata&1&75&40&2&-13.00&28\\
MCG-2-8-14&72.5&2&u&0.20&103	$\pm$	30&\nodata&1&\nodata&1&\nodata&1&\nodata&1&\nodata&95&2&\nodata&1\\
MCG-2-8-39&133.0&2&n&0.11&58	$\pm$	16&\nodata&1&\nodata&1&\nodata&1&\nodata&1&\nodata&3&2&-12.84&33\\
\rowstyle{\bfseries} MCG-3-34-64&79.3&1.8/2&y&0.69&51	$\pm$	8&\nodata&1&39&51&109&52&\nodata&1&29	$\pm$	14&75&2&-11.96&28\\
MCG-5-23-16&42.8&1.9&n&\nodata&76	$\pm$	27&40&53&\nodata&1&20&40&\nodata&1&40&52&2&-12.55&28\\
MCG-6-30-15&38.8&1.5&p&0.16&106	$\pm$	13&115&53, 29&\nodata&1&\nodata&1&\nodata&1&115&116&2&-12.63&28\\
MR 2251-178&293.0&1.5&u&0.51&57	$\pm$	23&\nodata&1&\nodata&1&\nodata&1&\nodata&1&\nodata&75&2&\nodata&1\\
Mrk 3&60.6&2&p&0.38&69&70&54, 55&84&56&150&57&\nodata&1&71	$\pm$	12&15&2&-11.71&11\\
Mrk 304&301.0&1&p&1.06&3&\nodata&1&\nodata&1&\nodata&1&\nodata&1&\nodata&\nodata&1&-13.90&58\\
Mrk 509&153.0&1.5&n&\nodata&107	$\pm$	24&\nodata&1&\nodata&1&\nodata&1&\nodata&1&\nodata&80&2&-12.55&19\\
Mrk 573&73.1&2&u&&132	$\pm$	21&124&59, 55&125&60&51&61&\nodata&1&130	$\pm$	9&180&30&-12.10&33\\
Mrk 590&116.0&1&n&0.02&133	$\pm$	24&175&29&\nodata&1&\nodata&1&\nodata&1&175&\nodata&1&-13.51&11\\
Mrk 841&170.0&1.5&p&0.42&155	$\pm$	31&\nodata&1&\nodata&1&\nodata&1&\nodata&1&\nodata&70&2&-12.62&33\\
Mrk 915&104.0&1.9&u&0.19&14&5&29&\nodata&1&\nodata&1&\nodata&1&5&-9&2&-12.41&34\\
Mrk 926&210.0&1.5&p&0.56&66	$\pm$	15&90&62&90&63&\nodata&1&\nodata&1&90&104&2&-12.90&11\\
Mrk 1014&807.0&1.5&p&0.81&63&90&64&90&65&\nodata&1&\nodata&1&90&74&2&-12.92&11\\
Mrk 1018&191.0&1&n&\nodata&95	$\pm$	15&\nodata&1&\nodata&1&\nodata&1&\nodata&1&\nodata&176&2&-13.74&34\\
Mrk 1239&95.4&1n&p&0.12&122	$\pm$	40&191&35&\nodata&1&\nodata&1&\nodata&1&191&\nodata&1&-12.81&15\\
NGC 235A&96.2&2&p&0.67&97	$\pm$	40&80&35&43&66&\nodata&1&\nodata&1&62	$\pm$	26&120&2&-12.64&34\\
NGC 424&49.5&2&p&0.45&80&70&67&90&68, 32&43&69&\nodata&1&68	$\pm$	24&61&2&-12.59&33\\
NGC 454E&52.3&2&p&0.25&121	$\pm$	5&\nodata&1&\nodata&1&\nodata&1&\nodata&1&\nodata&90&2&-12.80&28\\
NGC 526A&82.8&1.9&n&\nodata&132	$\pm$	13&130&35, 70&\nodata&1&\nodata&1&\nodata&1&130&110&2&-12.71&15\\
NGC 788&57.2&2&n&0.22&38	$\pm$	34&\nodata&1&62&66&124&71&\nodata&1&48	$\pm$	20&108&2&-12.74&19\\
NGC 985&195.0&1.5&n&\nodata&87	$\pm$	21&\nodata&1&\nodata&1&\nodata&1&\nodata&1&\nodata&80&2&-12.66&34\\
\rowstyle{\bfseries} NGC 1068&14.4&1.8/2&y&1.74&175	$\pm$	3&215&72, 73&201&74&140&75&94&76, 77, 78&208	$\pm$	20&245&30&-10.72&19\\
NGC 1144&128.0&2&u&\nodata&105	$\pm$	25&\nodata&1&\nodata&1&\nodata&1&\nodata&1&\nodata&118&2&-13.27&15\\
NGC 1194&58.2&1.9&n&\nodata&102	$\pm$	40&56&29&56&51&\nodata&1&\nodata&1&56&139&2&-12.84&28\\
NGC 1275&76.8&1.5/L&u&0.02&210	$\pm$	16&\nodata&1&160&79, 80, 81&\nodata&1&\nodata&1&160&290&2&$\le$	-12.73&19\\
NGC 1365&17.9&1.8&u&0.12&82	$\pm$	25&125&82, 83, 84, 85&125&86&\nodata&1&\nodata&1&125&20&2&-11.80&19\\
\rowstyle{\bfseries} NGC 1386&16.5&2&y&0.79&177	$\pm$	20&185&29&170&66&\nodata&1&\nodata&1&178	$\pm$	11&205&2&-12.06&19\\
NGC 1566&14.3&1.5&u&0.06&157	$\pm$	37&135&55&\nodata&1&\nodata&1&\nodata&1&135&225&2&-13.05&19\\
NGC 1667&67.8&2&p&0.32&54&\nodata&1&\nodata&1&\nodata&1&\nodata&1&\nodata&3&2&-13.03&19\\
NGC 2110&35.9&2&p&0.12&115	$\pm$	38&160&87&180&88, 66&83&89&\nodata&1&171	$\pm$	10&175&2&-12.34&28\\
\rowstyle{\bfseries} NGC 2992&39.7&1.5/2&y&\nodata&38	$\pm$	43&125&90&154&91, 92&35&40&\nodata&1&-41	$\pm$	21&17&2&-11.97&19\\
\rowstyle{\bfseries} NGC 3081&40.9&2&p&0.21&170	$\pm$	10&164&53&164&66&79&69&\nodata&1&166	$\pm$	3&251&2&-12.00&19\\
NGC 3147&30.1&2&p&0.33&152	$\pm$	5&\nodata&1&\nodata&1&\nodata&1&\nodata&1&\nodata&150&2&$\le$	-13.19&19\\
\rowstyle{\bfseries} NGC 3227&22.1&1.5&y&0.59&158	$\pm$	40&195&55&170&93&\nodata&1&\nodata&1&183	$\pm$	18&157&2&-12.24&19\\
\rowstyle{\bfseries} NGC 3281&52.8&2&y&0.34&176	$\pm$	16&200&29&\nodata&1&\nodata&1&\nodata&1&200&140&2&-11.86&19\\
NGC 3393&61.6&2&u&0.29&81	$\pm$	57&55&55, 94&50&32, 51&\nodata&1&-34&95, 96&65	$\pm$	22&13&2&\nodata&1\\
NGC 3783&48.4&1.5&u&0.12&122	$\pm$	35&\nodata&1&\nodata&1&-45&97&\nodata&1&\nodata&163&2&-12.55&19\\
NGC 3982&21.4&2&u&0.20&37	$\pm$	39&\nodata&1&\nodata&1&\nodata&1&\nodata&1&\nodata&25&2&$\le$	-12.93&19\\
NGC 4051&12.2&1n&u&0.06&134&100&55&\nodata&1&\nodata&1&\nodata&1&100&130&30&-12.58&19\\
NGC 4074&107.0&2&p&0.43&189&131&35&133&66&\nodata&1&\nodata&1&132	$\pm$	1&278&2&-12.55&34\\
NGC 4138&13.8&1.9&u&0.25&147&\nodata&1&\nodata&1&\nodata&1&\nodata&1&\nodata&150&2&-13.37&19\\
NGC 4151&13.3&1.5&p&0.28&37	$\pm$	36&67&98, 99, 100&77&101&\nodata&1&\nodata&1&72	$\pm$	7&-50&30&-11.68&19\\
NGC 4235&41.2&1.2&n&\nodata&66	$\pm$	61&48&98&\nodata&1&\nodata&1&\nodata&1&48&49&2&-13.36&19\\
NGC 4258&7.6&2&u&0.04&1	$\pm$	51&\nodata&1&\nodata&1&\nodata&1&83&102&173&150&2&-13.13&19\\
\rowstyle{\bfseries} NGC 4388&19.2&2&y&0.62&28	$\pm$	34&30&103&30&103&\nodata&1&\nodata&1&30&91&2&-11.59&19\\
NGC 4395&4.3&1.8&u&0.64&101&90&104, 105&\nodata&1&\nodata&1&\nodata&1&90&127&2&-13.37&19\\
NGC 4501&17.9&2&u&\nodata&153&\nodata&1&\nodata&1&\nodata&1&\nodata&1&\nodata&138&2&-13.40&19\\
NGC 4507&57.5&2&n&\nodata&111	$\pm$	22&145&29&190&32&37&69&\nodata&1&154	$\pm$	32&50&2&-12.48&19\\
\rowstyle{\bfseries} NGC 4593&45.6&1&y&0.18&103	$\pm$	19&100&29&\nodata&1&\nodata&1&\nodata&1&100&50&30&-12.88&19\\
NGC 4941&21.2&2&n&0.01&116	$\pm$	11&\nodata&1&155&51&\nodata&1&\nodata&1&155&195&30&-12.82&19\\
NGC 4992&119.0&1/2/L/N&u&0.44&127	$\pm$	53&\nodata&1&\nodata&1&\nodata&1&\nodata&1&\nodata&192&2&-13.96&34\\
\rowstyle{\bfseries} NGC 5033&18.1&1.2&y&0.11&175	$\pm$	5&80&106&95&48&\nodata&1&\nodata&1&88	$\pm$	11&172&2&-12.80&19\\
NGC 5135&66.0&2&n&\nodata&66	$\pm$	48&0&107&30&68&\nodata&1&\nodata&1&15	$\pm$	21&120&30&-12.23&19\\
NGC 5252&109.0&1.9&u&0.08&144	$\pm$	24&167&108&175&109&\nodata&1&\nodata&1&171	$\pm$	6&192&2&-13.06&34\\
NGC 5273&15.3&1.5&p&0.05&115	$\pm$	44&180&53&185&91, 66&\nodata&1&\nodata&1&183	$\pm$	4&184&2&-13.43&19\\
NGC 5347&38.1&2&u&\nodata&3	$\pm$	5&30&98, 29&\nodata&1&\nodata&1&\nodata&1&30&-60&2&-13.12&33\\
NGC 5506&31.6&2&y&0.19&60	$\pm$	56&-16&110&\nodata&1&80&40&\nodata&1&-13	$\pm$	4&89&2&-11.65&19\\
NGC 5548&80.7&1.5&p&1.18&200	$\pm$	17&180&111, 29&160&112, 48&\nodata&1&\nodata&1&170	$\pm$	14&123&2&-12.90&11\\
NGC 5643&20.9&2&p&0.37&52	$\pm$	17&84&113&87&114, 115&\nodata&1&\nodata&1&86	$\pm$	2&90&30&-12.09&19\\
\rowstyle{\bfseries} NGC 5728&45.4&1.9/2&y&0.41&98	$\pm$	25&120&116&127&117&\nodata&1&\nodata&1&124	$\pm$	5&35&30&-11.89&19\\
NGC 5995&117.0&1.9&n&0.02&111	$\pm$	39&\nodata&1&\nodata&1&\nodata&1&\nodata&1&\nodata&147&2&-12.89&15\\
NGC 6251&112.0&1/2/L&u&0.89&145&\nodata&1&116&118, 119&\nodata&1&\nodata&1&116&200&2&-14.28&120\\
NGC 6300&14.3&2&p&0.43&112	$\pm$	4&\nodata&1&190&32&35&40&\nodata&1&158	$\pm$	46&118&2&-12.53&19\\
NGC 6814&20.1&1.5&n&0.26&97	$\pm$	10&150&55&180&91&\nodata&1&\nodata&1&165	$\pm$	21&80&30&-12.67&19\\
NGC 6860&65.8&1.5&n&0.03&53	$\pm$	49&90&121, 29&\nodata&1&\nodata&1&\nodata&1&90&32&2&-12.92&15\\
NGC 6890&33.8&1.9/2&p&0.24&102&\nodata&1&\nodata&1&\nodata&1&\nodata&1&\nodata&75&30&-13.00&33\\
\rowstyle{\bfseries} NGC 7172&34.8&2&y&0.40&86	$\pm$	10&60&122&90&32&\nodata&1&\nodata&1&75	$\pm$	21&105&2&-12.31&19\\
NGC 7212&116.0&2&p&0.54&177&170&29, 123&170&103&100&124&\nodata&1&177	$\pm$	12&212&2&\nodata&1\\
NGC 7213&23.0&1.5/L&n&0.18&104	$\pm$	22&90&1&\nodata&1&\nodata&1&\nodata&1&90&121&2&-13.68&19\\
\rowstyle{\bfseries} NGC 7314&18.3&1.9/2&y&0.41&87	$\pm$	43&\nodata&1&\nodata&1&5&40&\nodata&1&93&3&2&-12.31&19\\
\rowstyle{\bfseries} NGC 7469&67.9&1/1.5&y&0.11&110	$\pm$	32&90&125&85&126&\nodata&1&\nodata&1&88	$\pm$	4&127&2&-12.44&19\\
NGC 7479&30.0&2&n&\nodata&82&80&127&\nodata&1&\nodata&1&\nodata&1&80&0&30&$\le$	-12.57&19\\
\rowstyle{\bfseries} NGC 7674&126.0&2&y&0.26&140	$\pm$	24&120&29&115&128&40&57&\nodata&1&122	$\pm$	8&165&2&-12.35&45\\
PG 0026+129&691.0&1.2&u&0.16&107&\nodata&1&\nodata&1&\nodata&1&\nodata&1&\nodata&\nodata&1&-13.70&11\\
PG 0052+251&759.0&1.2&u&\nodata&129&\nodata&1&\nodata&1&\nodata&1&\nodata&1&\nodata&\nodata&1&\nodata&1\\
PG 0844+349&302.0&1&p&0.74&109&\nodata&1&\nodata&1&\nodata&1&\nodata&1&\nodata&50&2&-13.83&45\\
PG 2130+099&288.0&1.5&u&0.16&158	$\pm$	27&\nodata&1&135&1&\nodata&1&\nodata&1&135&242&2&-13.00&45\\
PKS 1417-19&586.0&1.5&p&0.65&98&\nodata&1&20&129&\nodata&1&\nodata&1&20&40&2&\nodata&1\\
PKS 1814-63&302.0&2&u&0.27&25&\nodata&1&\nodata&1&\nodata&1&\nodata&1&\nodata&114&2&-13.49&9\\
PKS 2158-380&149.0&2&u&0.10&109&\nodata&1&90&130, 131&\nodata&1&\nodata&1&90&103&2&\nodata&1\\
Pictor A&161.0&1.5/L&u&\nodata&99	$\pm$	21&135&132&101&133, 134&\nodata&1&\nodata&1&118	$\pm$	24&82&2&$\le$	-14.30&34\\
Superantennae S&291.0&2&p&0.39&151	$\pm$	26&\nodata&1&\nodata&1&\nodata&1&\nodata&1&\nodata&175&2&-13.20&33\\
UGC 12348&110.0&2&p&0.13&72	$\pm$	3&100&29&\nodata&1&\nodata&1&\nodata&1&100&136&2&\nodata&1\\
Z 41-20&170.0&2&u&\nodata&172	$\pm$	36&\nodata&1&\nodata&1&\nodata&1&\nodata&1&\nodata&215&2&\nodata&1\\

\enddata						                              
\tablecomments{
Objects in bold belong to the extended Seyfert sample.
(1), (2), (3), and (4) short object name, distance, optical class, nuclear MIR extension (``n'' not resolved, ``p'' possibly extended, ``u'' unknown extension,  ``y'' extended) from \cite{asmus_subarcsecond_2014};
(5) relative extended to nuclear emission ratio at $12\,\micron$, $\extnuc = \Fgau / \Fpsf - 1$; 
(6) mean nuclear MIR PA from the individual images from \cite{asmus_subarcsecond_2014} with standard deviation in case of multiple measurements available;
note that this value can be unreliable if the object was not classified as extended; 
(7) PA of the NLR major axis, mostly from \oiii images;
(8) reference for the NLR PA;
(9) PA of the nuclear radio morphology;
(10) reference for the radio PA;
(11) PA of polarised broad lines;
(12) reference for the polarisation PA;
(13) maser PA;
(14) reference for the maser PA;
(15) mean system axis PA and standard deviation (if multiple PAs are used);
(16) host inner structure PA;
(17) host PA reference;
(18) \oiv flux;
(19) \oiv reference; 
list of references:
1: Hyperleda \citep{makarov_hyperleda._2014}; 
2: \cite{leahy_vla_1991}; 
3: \cite{giovannini_bologna_2005}; 
4: \cite{cohen_polarimetry_1999}; 
5: \cite{ogle_spitzer_2006}; 
6: \cite{baum_extended_1988}; 
7: \cite{leahy_study_1997}; 
8: \cite{dicken_spitzer_2014}; 
9: \cite{walker_radio_1987}; 
10: \cite{dasyra_view_2011}; 
11: \cite{black_study_1992}; 
12: \cite{young_scattered_1996}; 
13: \cite{leahy_jets_1995}; 
14: \cite{tommasin_spitzer-irs_2010}; 
15: \cite{mayer_multifrequency_1979}; 
16: \cite{morganti_nature_1991}; 
17: \cite{schreier_detection_1981}; 
18: \cite{diamond-stanic_isotropic_2009}; 
19: \cite{marconi_prominent_1994}; 
20: \cite{wilson_hubble_2000}; 
21: \cite{oliva_spectropolarimetry_1998}; 
22: \cite{greenhill_warped_2003}; 
23: \cite{jackson_cygnus_1998}; 
24: \cite{canalizo_adaptive_2003}; 
25: \cite{perley_jet_1984}; 
26: \cite{privon_modeling_2012}; 
27: \cite{weaver_mid-infrared_2010}; 
28: \cite{schmitt_hubble_2003}; 
29: this work; 
30: \cite{schmitt_anisotropic_1995}; 
31: \cite{morganti_radio_1999}; 
32: \cite{wu_diagnostics_2011}; 
33: \cite{weedman_infrared_2012}; 
34: \cite{mulchaey_emission-line_1996}; 
35: \cite{schmid_spectropolarimetry_2003}; 
36: \cite{erwin_double-barred_2004}; 
37: \cite{fraquelli_extended_2000}; 
38: \cite{falcke_helical_1996}; 
39: \cite{lumsden_spectropolarimetry_2004}; 
40: \cite{pereira-santaella_mid-infrared_2010}; 
41: \cite{morganti_radio_1998}; 
42: \cite{farrah_high-resolution_2007}; 
43: \cite{young_polarimetry_1996}; 
44: \cite{dasyra_high-ionization_2008}; 
45: \cite{willett_mid-infrared_2011}; 
46: \cite{ford_bubbles_1985}; 
47: \cite{ho_radio_2001}; 
48: \cite{hagiwara_water_2001}; 
49: \cite{hagiwara_low-luminosity_2007}; 
50: \cite{schmitt_jet_2001}; 
51: \cite{kay_blue_1994}; 
52: \cite{ferruit_hubble_2000}; 
53: \cite{capetti_morphology_1995}; 
54: \cite{schmitt_comparison_1996}; 
55: \cite{kukula_high-resolution_1993}; 
56: \cite{miller_spectropolarimetry_1990}; 
57: \cite{schweitzer_spitzer_2006}; 
58: \cite{pogge_imaging_1995}; 
59: \cite{ulvestad_radio_1984-1}; 
60: \cite{nagao_detection_2004}; 
61: \cite{durret_long_1988}; 
62: \cite{mundell_parsec-scale_2000}; 
63: \cite{bennert_size_2002}; 
64: \cite{leipski_radio_2006}; 
65: \cite{nagar_radio_1999}; 
66: \cite{durret_imaging_1987}; 
67: \cite{ulvestad_radio_1989}; 
68: \cite{moran_frequency_2000}; 
69: \cite{bennert_size_2006}; 
70: \cite{kay_hidden_1998}; 
71: \cite{pogge_extended_1988}; 
72: \cite{evans_hst_1991}; 
73: \cite{gallimore_subarcsecond_1996-1}; 
74: \cite{bailey_polarization_1988}; 
75: \cite{claussen_water-vapor_1984}; 
76: \cite{greenhill_vlbi_1996}; 
77: \cite{gallimore_h_1996}; 
78: \cite{pedlar_radio_1990}; 
79: \cite{walker_detection_1994}; 
80: \cite{vermeulen_discovery_1994}; 
81: \cite{phillips_remarkable_1983}; 
82: \cite{jorsater_kinematics_1984}; 
83: \cite{storchi-bergmann_detection_1991}; 
84: \cite{kristen_imaging_1997}; 
85: \cite{sandqvist_central_1995}; 
86: \cite{mulchaey_hubble_1994}; 
87: \cite{ulvestad_nuclear_1983}; 
88: \cite{moran_transient_2007}; 
89: \cite{allen_physical_1999}; 
90: \cite{ulvestad_radio_1984}; 
91: \cite{wehrle_radio_1988}; 
92: \cite{mundell_anisotropic_1995}; 
93: \cite{cooke_narrow-line_2000}; 
94: \cite{kondratko_discovery_2006}; 
95: \cite{kondratko_parsec-scale_2008}; 
96: \cite{smith_seyferts_2004}; 
97: \cite{pogge_circumnuclear_1989}; 
98: \cite{perez_complex_1989}; 
99: \cite{evans_hubble_1993}; 
100: \cite{wilson_radio_1982}; 
101: \cite{greenhill_detection_1995}; 
102: \cite{falcke_hst_1998}; 
103: \cite{matthews_wfpc2_1999}; 
104: \cite{filippenko_low-mass_2003}; 
105: \cite{mediavilla_asymmetrical_2005}; 
106: \cite{gonzalez_delgado_ultraviolet-optical_1998}; 
107: \cite{tadhunter_anisotropic_1989}; 
108: \cite{wilson_ionization_1994}; 
109: \cite{colbert_large-scale_1996}; 
110: \cite{wilson_kinematics_1989}; 
111: \cite{wilson_radio_1982-1}; 
112: \cite{simpson_one-sided_1997}; 
113: \cite{morris_velocity_1985}; 
114: \cite{ leipski_radio_2006}; 
115: \cite{wilson_ionization_1993}; 
116: \cite{schommer_ionized_1988}; 
117: \cite{perley_high-resolution_1984}; 
118: \cite{jones_high_1986}; 
119: \cite{melendez_new_2008}; 
120: \cite{lipari_high-resolution_1993}; 
121: \cite{smajic_unveiling_2012}; 
122: \cite{cracco_origin_2011}; 
123: \cite{tran_nature_1995}; 
124: \cite{muller-sanchez_outflows_2011}; 
125: \cite{lonsdale_vlbi_2003}; 
126: \cite{yankulova_luminous_2007}; 
127: \cite{momjian_sensitive_2003}; 
128: \cite{antonucci_vla_1985}; 
129: \cite{boyce_faint_1996}; 
130: \cite{zirbel_ultraviolet_1998}; 
131: \cite{simkin_pictor_1999}; 
132: \cite{perley_radio_1997}. 
}
\end{deluxetable}
\clearpage
\end{landscape}

}

%\appendix

%\subsection{Notes on selected objects}\label{app:obj}

%\subsubsection{3C\,264 -- NGC\,3862}

\end{document}